\documentclass[aps,prc,floatfix,nofootinbib,showpacs,twocolumn,superscriptaddress]{revtex4-1}

\usepackage{mathrsfs}
\usepackage{amsmath}
\usepackage{amssymb}
\usepackage{color}
\usepackage{graphicx}
\usepackage{pzccal}
\usepackage{soul}

\begin{document}

\vspace*{-5.8ex}
\hspace*{\fill}{NPAC-12-11}

\vspace*{+4.8ex}

\title{Electric dipole moment of the $\rho$-meson}

\author{Mario Pitschmann}
\affiliation{University of Wisconsin-Madison, Madison, Wisconsin 53706, USA}
\affiliation{Physics Division, Argonne National Laboratory, Argonne, Illinois 60439, USA}

\author{Chien-Yeah Seng}
\affiliation{University of Wisconsin-Madison, Madison, Wisconsin 53706, USA}

\author{Michael J.\ Ramsey-Musolf}
\affiliation{University of Wisconsin-Madison, Madison, Wisconsin 53706, USA}
\affiliation{California Institute of Technology, Pasadena, California 91125, USA}

\author{Craig D.\ Roberts}
\affiliation{Physics Division, Argonne National Laboratory, Argonne, Illinois 60439, USA}
\affiliation{Department of Physics, Illinois Institute of Technology, Chicago, Illinois}

\author{Sebastian M.~Schmidt}
\affiliation{Institute for Advanced Simulation, Forschungszentrum J\"ulich and JARA, D-52425 J\"ulich, Germany}
%\affiliation{RWTH Aachen, I.\ Physikalisches Institut, Sommerfeldstr.\ 14,D-52074 Aachen, Germany}

\author{David J.~Wilson}
\affiliation{Physics Division, Argonne National Laboratory, Argonne, Illinois 60439, USA}

\date{19 September 2012}
%\date{29 August 2012}
%\date{01 July 2012}
%\date{19 May 2012}
%\date{24 January 2012}

\begin{abstract}
At an hadronic scale the effect of CP-violating interactions that typically appear in extensions of the Standard Model may be described by an effective Lagrangian, in which the operators are expressed in terms of lepton and partonic gluon and quark fields, and ordered by their mass dimension, $k\geq 4$.
Using a global-symmetry-preserving truncation of QCD's Dyson-Schwinger equations, we compute the $\rho$-meson's electric dipole moment (EDM), $d_\rho$, as generated by the leading dimension-four and -five CP-violating operators and an example of a dimension-six four-quark operator.
The two dimension-five operators; viz., quark-EDM and -chromo-EDM, produce contributions to $d_\rho$ whose coefficients are of the same sign and within a factor of two in magnitude.
Moreover, should a suppression mechanism be verified for the $\theta$-term in any beyond-Standard-Model theory, the contribution from a four-quark operator can match the quark-EDM and -chromo-EDM in importance.
This study serves as a prototype for the more challenging task of computing the neutron's EDM.
\end{abstract}

%\pacs{72.10.Bg, 73.40.Gk, 73.63.Kv, 71.10.Pm}

\maketitle

\section{Introduction}
%\noindent \emph{\textbf{Background}}.
The action for any local quantum field theory is invariant under the transformation generated by the antiunitary operator $CPT$, which is the product of the inversions: $C$, charge conjugation; $P$, parity transformation; and $T$, time reversal.  The combined $CPT$ transformation provides a rigorous correspondence between particles and antiparticles, and it relates the $S$ matrix for any given process to its inverse, where all spins are flipped and the particles replaced by their antiparticles.  Lorentz and $CPT$ symmetry together have many consequences, amongst them, that the mass and total width of any particle are identical to those of its antiparticle.

It is within this context that the search for the intrinsic electric dipole moment (EDM) of an elementary or composite but fundamental particle has held the fascination of physicists for over sixty years \cite{Purcell:1950zz}.  Its existence indicates the simultaneous violation of parity- and time-reversal-invariance in the theory that describes the particle's structure and interactions; and the violation of $P$- and $T$-invariance entails that $CP$ symmetry is also broken.
%Under the reflection of spatial coordinates, P (B · S) = B · S, whereas P (E · S) = -E · S. The presence of a non-zero d would therefore signify the existence of parity violation. It was soon realized that d also breaks time-reversal invariance. Indeed, under time reflection, T (B · S) = B · S and T (E · S) = -E · S. Therefore a non-zero d may exist if and only if both parity and time-reversal invariance are broken.
%
This last is critical for our existence because we represent a macroscopic excess of matter over antimatter.  As first observed by Sakharov \cite{Sakharov:1967dj}, in order for a theory to explain an excess of baryon matter, it must include processes that change baryon number, and break $C$- and $CP$-symmetries; and the relevant processes must have taken place out of equilibrium, otherwise they would merely have balanced matter and antimatter. (Alternately, the presence of CPT violation can circumvent the out-of-equilibrium environment.)

The electroweak component of the Standard Model (SM) is capable of satisfying Sakharov's conditions, owing to the existence of a complex phase in the $3\times 3$-CKM matrix which enables processes that mix all three quark generations.  However, this high-order process is too weak to explain the observed matter-antimatter asymmetry \cite{Cohen:1993nk,Trodden:1998ym,Morrissey:2012db}.  Hence, it is widely expected that any description of baryogenesis will require new sources of CP violation beyond the SM.
%even with the most precise measurement techniques available nowadays \cite{Harris,Lamoreaux,Pospelov,LEPTSUSY}.
This presents little difficulty, however, because extensions of the SM typically possess $CP$-violating interactions, whose parameters must, in fact, be tuned to small values in order to avoid conflict with known bounds on the size of such EDMs \cite{Harris:1999jx,Pospelov:2005pr,RamseyMusolf:2006vr,Lamoreaux:2009zz,Morrissey:2012db}. (For recent analyses, see, e.g., Refs.\,\cite{Cirigliano:2009yd,Li:2010ax,Kozaczuk:2012xv} and references therein.)

%Therefore, the detection of an EDM necessarily implies new physics beyond standard model. On the other hand, free parameters of various Beyond-Standard-Model (BSM) theories (e.g. supersymmetry (SUSY)) can be constrained by lowering the bounds of the EDM.
%%%
%The connection between the observed EDM and parameters in CP-violating high energy theories is an important issue. Theories BSM such as SUSY may provide CP-odd phases at high energy (e.g. the TeV scale). However in low-energy precision measurements one deals with nucleons or atoms at an energy scale of $\Lambda_{\mathrm{QCD}}\sim 240\mathrm{MeV}$. For that aim, the high-energy degrees of freedom have to be integrated out to obtain a low-energy effective theory of quarks and gluons. Next, atomic or nuclear many-body calculations have to be performed in order to match the Wilson coefficients of the effective theories to the observable EDM. This calculation involves non-perturbative QCD rendering first-principle calculations impossible. Nevertheless, there are several models containing different kinds of approximations such as Schwinger-Dyson equations~\cite{Hecht}, Chiral Perturbation Theory~\cite{Ji}, naive quark models and so on.

The question here is how such bounds should be imposed.  That is not a problem for elementary particles, like the electron.  However, it is a challenge when the SM extension produces an operator involving current-quarks and/or gluons.  In that case the $CP$ violation is expressed as an hadronic property and one must have at hand a nonperturbative method with which to compute the impact of $CP$-violating features of partonic quarks and gluons on the hadronic composite.

To elucidate, extensions of the SM are typically active at some large but unspecified energy-scale, $\Lambda$, and their effect at an hadronic scale is expressed in a low-energy effective Lagrangian:
%%%http://arxiv.org/pdf/hep-ph/9910273v1.pdf
\begin{equation}
{\cal L}_{\rm eff} \sim \sum_{j,k} K_j \, {\cal O}_j^{(k)} \Lambda^{4-k},
\label{Leff}
\end{equation}
where ${\cal O}_j^{(k)}$ are composite $CP$-odd local operators of dimension $k\geq 4$ and $\{K_j\}$ are dimensionless strength parameters, which monitor the size of the model's $CP$-violating phases and commonly evolve logarithmically with the energy scale.  The calculation of an hadronic EDM therefore proceeds in two steps.  The first, easier, part requires calculation of the coefficients $\{K_i\}$ in a given model.  This involves the systematic elimination of degrees-of-freedom that are irrelevant at energy-scales less than $\Lambda$.  The second, far more challenging exercise, is the nonperturbative problem of translating the current-quark-level interaction in Eq.\,\eqref{Leff} into observable properties of hadrons.

We illustrate the procedure in the case of the $\rho$-meson.  Not that there is any hope of measuring a $\rho$-meson EDM but because the nonperturbative methods necessary can most readily be illustrated in the case of systems defined by two valence-quark degrees-of-freedom.  In taking this path, we follow other authors \cite{Hecht:1997uj,Hecht:1999fd,Pospelov:1999rg} but will nonetheless expose novel insights, especially because we consider more operator structures than have previously been considered within a single unifying framework.
It is worth remarking here that particles with spin also possess a magnetic dipole moment. That moment is aligned with the particle's spin because it is the only vector available.  The same is true of the expectation value of any electric dipole moment.

Herein we shall estimate the contribution of some dimension four, five and six operators to the EDM of the $\rho^+$-meson; viz., the impact on the $\rho$ of the local Lagrangian density
\begin{eqnarray}
\nonumber
{\cal L}_{\rm eff}
&=&-i\bar \theta \frac{g_s^2}{32\pi^2} \, G^a_{\mu\nu} \tilde G^a_{\mu\nu}
- \frac{i}{2} \sum_{q=u,d} d_q \, \bar q \, \gamma_5 \sigma_{\mu\nu}  q \, F_{\mu\nu}\\
\nonumber
&&  - \frac{i}{2} \sum_{q=u,d} \tilde d_q \, \bar q \,  \mbox{\small $\frac{1}{2}$}
\lambda^a \gamma_5\sigma_{\mu\nu}  q \, g_s G^a_{\mu\nu}\\
&& + \frac{\mathpzc{K}}{\Lambda^2} \, i\varepsilon_{jk} \left[
    \bar Q_j d \, \bar Q_k \gamma_5 u + {\rm h.c.} \right],
    \label{LeffCompute}
\end{eqnarray}
where: latin superscripts represent colour; $g_s$ is the strong coupling constant; $F_{\mu\nu}$ and $G_{\mu\nu}^a$ are photon and gluon field-strength tensors, respectively, and $\tilde G_{\mu\nu}^a = (1/2) \epsilon_{\mu\nu\lambda\rho}G_{\lambda\rho}^a$; $\{\bar Q_i| i=1,2 \} = \{ \bar u_L,\bar d_L \}$, with the subscript indicating left-handed; $\bar\theta$ is QCD's effective $\theta$-parameter, which combines $\theta_{\rm QCD}$ and the unknown phase of the current-quark-mass matrix; and $\{d_q \}$, $\{\tilde d_q\}$ are quark EDMs and chromo-EDMs, respectively.

We note that Eq.\,\eqref{LeffCompute} is expressed at a renormalisation scale $\zeta \sim 2\,$GeV, which is far below that of electroweak symmetry breaking but still within the domain upon which perturbative QCD is applicable.  Moreover, we have chosen to include just one dimension-six operator in the Lagrangian; i.e., a particular type of four-fermion interaction.  There is a host of dimension-six operators, Weinberg's CP-odd three-gluon vertex amongst them \cite{Weinberg:1989dx}.  However, for our illustrative purpose, nothing is lost by omitting them because the potency of the one operator we do consider can serve as an indication of the strength with which each might contribute.

One merit of our analysis of the contribution from Eq.\,\eqref{LeffCompute} to the EDM of the $\rho^+$-meson is the connection of these EDM responses with values of a vast array of hadron observables that are all computed within precisely the same framework using exactly the same parameters \cite{GutierrezGuerrero:2010md,Roberts:2010rn,Roberts:2011cf,%
Roberts:2011wy,Wilson:2011aa,Chen:2012qr}.  We explain this framework in Sec.\,\ref{rhomodel}.  In addition to providing the first such comprehensive treatment, our study is novel in considering the impact of a dimension-six operator on the $\rho^+$-meson's EDM.

We introduce the $\rho$-meson electromagnetic form factors in Sec.\,\ref{sec:formfactors}.  The effects of Eq.\,\eqref{LeffCompute} on the $\rho$-meson bound-state are analysed in Sec.\,\ref{sec:formulae}.  Each interaction term is considered separately, so that we present a raft of algebraic formulae that are readily combined, evaluated and interpreted.  Numerical results are provided in Sec.\,\ref{sec:results} and placed in context with previous studies.  Section~\ref{sec:epilogue} is an epilogue.

\section{$\mathbf{\rho}$-meson as a Bound State}
\label{rhomodel}
\subsection{$\mathbf{\rho}$-$\mathbf{\gamma}$ Vertex}
The $\rho^+$-meson is a composite particle and thus its EDM appears in the dressed vertex that describes its coupling with the photon; viz.,
\begin{eqnarray}
\nonumber\lefteqn{
\mathcal P_{\alpha{\alpha'}}^T(p)\Gamma_{{\alpha'}\mu{\beta'}}(p,p^\prime)\mathcal P_{{\beta'}\beta}^T(p')}\\
\nonumber
&=& \mathcal P_{\alpha{\alpha'}}^T(p)\Big\{
(p + p')_\mu[-\delta_{{\alpha'}{\beta'}}\mathcal E(q^2) + q_{\alpha'} q_{\beta'}\mathcal Q(q^2)]\\
\nonumber
&& \quad  -(\delta_{\mu{\alpha'}}q_{\beta'} - \delta_{\mu{\beta'}}q_{\alpha'})\mathcal M(q^2)\\
&& \quad - i\varepsilon_{{\alpha'}{\beta'}\mu\sigma}q_\sigma\mathcal D(q^2)\Big\}\,\mathcal P_{{\beta'}\beta}^T(p')\,,
\label{rgrvertex}
\end{eqnarray}
where: $p_\alpha$ is the momentum of the incoming $\rho$-meson; $p_\beta^\prime$, that of the outgoing $\rho$; $q_\mu=p_\mu^\prime - p_\mu$; and %the projector is
\begin{align}
  \mathcal P_{\alpha\beta}^T(p)&=\delta_{\alpha\beta} - \frac{p_\alpha p_\beta}{p^2}\,.
\end{align}
The vertex involves four scalar form factors whose $q^2=0$ values are understood as follows: $\mathcal E(0)$, electric charge, which is ``1'' in this case; $\mathcal M(0)$, magnetic moment, $\mu_\rho$, in units of $e/[2m_\rho]$, where $e$ is the magnitude of the electron charge; $\mathcal Q(0)=(2/m_\rho^2)(Q_\rho+\mu_\rho -1)$, with $Q_\rho$ the meson's electric quadrupole moment; and $\mathcal D(0)$ is the meson's electric dipole moment, in
units of $e/[2m_\rho]$. %which is customarily quoted in units of ``$e\,$cm''.

\subsection{Contact Interaction}
Our goal is calculation of the last of these,  $\mathcal D(0)$, and for this we choose to work within the continuum framework provided by QCD's Dyson-Schwinger equations (DSEs) \cite{Chang:2011vu,Bashir:2012fs,Roberts:2012sv}.  To be specific, we perform the computation using a global-symmetry-preserving treatment of a vector$\times$vector contact-interaction because that has proven to be a reliable explanatory and predictive tool for hadron properties measured with probe momenta less-than the dressed-quark mass, $M\sim 0.4\,$GeV \cite{GutierrezGuerrero:2010md,Roberts:2010rn,Roberts:2011cf,%
Roberts:2011wy,Wilson:2011aa,Chen:2012qr}.

To expand upon the reasons for this choice of interaction we note that DSE kernels with a closer connection to perturbative QCD; namely, which preserve QCD's one-loop renormalisation group behaviour, have long been employed in studies of the spectrum and interactions of mesons \cite{Jain:1993qh,Maris:1997tm,Maris:2003vk}.  Such kernels are developed in the rainbow-ladder approximation, which is the leading-order in a systematic and global-symmetry-preserving truncation scheme \cite{Munczek:1994zz,Bender:1996bb}; and their model input is expressed via a statement about the nature of the gap equation's kernel at infrared momenta.  With a single parameter that expresses a confinement length-scale or strength \cite{Maris:2002mt,Eichmann:2008ae}, they have successfully described and predicted numerous properties of vector \cite{Eichmann:2008ae,Maris:1999nt,Bhagwat:2006pu,Qin:2011dd,Qin:2011xq} and pseudoscalar mesons \cite{Eichmann:2008ae,Qin:2011dd,Qin:2011xq,Maris:1998hc,Maris:2000sk,Maris:2002mz,Bhagwat:2007ha} with masses less than 1\,GeV, and ground-state baryons \cite{Eichmann:2008ef,Eichmann:2011vu,Eichmann:2011ej,Eichmann:2011pv}.  Such kernels are also reliable for ground-state heavy-heavy mesons \cite{Bhagwat:2006xi}.  Given that contact-interaction results for low-energy observables are indistinguishable from those produced by the most sophisticated interactions, it is sensible to capitalise on the simplicity of the contact-interaction herein.

The starting point for our study is the dressed-quark propagator, which is obtained from the gap equation:
\begin{eqnarray}
\nonumber \lefteqn{S(p)^{-1}= i\gamma\cdot p + m}\\
&&+ \!\! \int \! \frac{d^4q}{(2\pi)^4} g^2 D_{\mu\nu}(p-q) \frac{\lambda^a}{2}\gamma_\mu S(q) \frac{\lambda^a}{2}\Gamma_\nu(q,p) ,\;
\label{gendse}
\end{eqnarray}
wherein $m$ is the Lagrangian current-quark mass, $D_{\mu\nu}$ is the vector-boson propagator and $\Gamma_\nu$ is the quark--vector-boson vertex.  We use
\begin{equation}
\label{njlgluon}
%g^2 D_{\mu \nu}(p-q) = \delta_{\mu \nu} \frac{1}{m_G^2}\,,
g^2 D_{\mu \nu}(p-q) = \delta_{\mu \nu} \frac{4 \pi \alpha_{\rm IR}}{m_G^2}\,,
\end{equation}
%where $m_G$ is a gluon mass-scale,
where $m_G=0.8\,$GeV is a gluon mass-scale typical of the one-loop renormalisation-group-improved interaction introduced in Ref.\,\cite{Qin:2011dd}, and the fitted parameter $\alpha_{\rm IR}/\pi = 0.93$ is commensurate with contemporary estimates of the zero-momentum value of a running-coupling in QCD \cite{Aguilar:2010gm,Boucaud:2010gr}.  Equation~\eqref{njlgluon} is embedded in a rainbow-ladder truncation of the DSEs, which is the leading-order in the most widely used, symmetry-preserving truncation scheme \cite{Bender:1996bb}.  This means
\begin{equation}
\label{RLvertex}
\Gamma_{\nu}(p,q) =\gamma_{\nu}
\end{equation}
in Eq.\,(\ref{gendse}) and in the subsequent construction of the Bethe-Salpeter kernels. One may view the interaction in Eq.\,(\ref{njlgluon}) as being inspired by models of the Nambu--Jona-Lasinio (NJL) type \cite{Nambu:1961tp}.  However, in implementing the interaction as an element in a rainbow-ladder truncation of the DSEs, our treatment is atypical; e.g., we have a single, unique coupling parameter, whereas common applications of the NJL model have different, tunable strength parameters for each collection of operators that mix under symmetry transformations.

Using Eqs.\,(\ref{njlgluon}), (\ref{RLvertex}), the gap equation becomes
\begin{equation}
 S^{-1}(p) =  i \gamma \cdot p + m +  \frac{16\pi}{3}\frac{\alpha_{\rm IR}}{m_G^2} \int\!\frac{d^4 q}{(2\pi)^4} \,
\gamma_{\mu} \, S(q) \, \gamma_{\mu}\,,   \label{gap-1}
\end{equation}
an equation in which the integral possesses a quadratic divergence, even in the chiral limit.  When the divergence is regularised in a Poincar\'e covariant manner, the solution is
\begin{equation}
\label{genS}
S(p)^{-1} = i \gamma\cdot p + M\,,
\end{equation}
where $M$ is momentum-independent and determined by
\begin{equation}
\label{gapM}
M = m + M\frac{4\alpha_{\rm IR}}{3\pi m_G^2} \int_0^\infty \!ds \, s\, \frac{1}{s+M^2}\,.
\end{equation}

Our regularisation procedure follows Ref.\,\cite{Ebert:1996vx}; i.e., we write
\begin{eqnarray}
\nonumber
\frac{1}{s+M^2} & = & \int_0^\infty d\tau\,{\rm e}^{-\tau (s+M^2)} \\
& \rightarrow & \int_{\tau_{\rm uv}^2}^{\tau_{\rm ir}^2} d\tau\,{\rm e}^{-\tau (s+M^2)}
\label{RegC}\\
& & =
\frac{{\rm e}^{- (s+M^2)\tau_{\rm uv}^2}-e^{-(s+M^2) \tau_{\rm ir}^2}}{s+M^2} \,, \label{ExplicitRS}
\end{eqnarray}
where $\tau_{\rm ir,uv}$ are, respectively, infrared and ultraviolet regulators.  It is apparent from Eq.\,(\ref{ExplicitRS}) that $\tau_{\rm ir}=:1/\Lambda_{\rm ir}$ finite implements confinement by ensuring the absence of quark production thresholds \cite{Krein:1990sf,Chang:2011vu}.  Since Eq.\,(\ref{njlgluon}) does not define a renormalisable theory, then $\Lambda_{\rm uv}:=1/\tau_{\rm uv}$ cannot be removed but instead plays a dynamical role, setting the scale of all dimensioned quantities.

Using Eq.\,\eqref{RegC}, the gap equation becomes
\begin{equation}
%M = m +  \frac{M}{3\pi^2 m_G^2} \,{\cal C}(M^2;\tau_{\rm ir},\tau_{\rm uv})\,,
%M = m +  \frac{M}{3\pi^2 m_G^2} \,{\cal C}^{\rm iu}(M^2)\,,
M = m + M\frac{4\alpha_{\rm IR}}{3\pi m_G^2}\,\,{\cal C}^{\rm iu}(M^2)\,,
\label{gapactual}
\end{equation}
where
\begin{eqnarray}
{\cal C}^{\rm iu}(M^2) &= & M^2 \overline{\cal C}^{\rm iu}(M^2)\\
 &=& M^2 \big[ \Gamma(-1,M^2 \tau_{\rm uv}^2) - \Gamma(-1,M^2 \tau_{\rm ir}^2)\big]\,,\rule{1ex}{0ex}
\end{eqnarray}
with $\Gamma(\alpha,y)$ the incomplete gamma-function, and, for later use, ${\cal C}_1^{\rm iu}(z) = -z(d/dz) {\cal C}^{\rm iu}(z)$.

In rainbow-ladder truncation, with the interaction in Eq.\,(\ref{njlgluon}), the homogeneous Bethe-Salpeter equation for the colour-singlet $\rho$-meson is
\begin{equation}
\Gamma_\mu^\rho(k;P) =
%-\frac{4}{3}\frac{1}{m_G^2}
-\frac{16 \pi}{3}\frac{\alpha_{\rm IR}}{m_G^2}
\int \! \frac{d^4q}{(2\pi)^4}\, \gamma_\sigma \chi_\mu^\rho(q;P)\gamma_\sigma \,,
\label{genbse}
\end{equation}
where $\chi_\mu^\rho(q;P) = S(q+P)\Gamma_\mu^\rho(q;P)S(q)$ and $\Gamma_\mu(q;P)$ is the meson's Bethe-Salpeter amplitude.  Since the integrand does not depend on the external relative-momentum, $k$, then a global-symmetry-preserving regularisation of Eq.\,(\ref{genbse}) yields solutions that are independent of $k$.  With a dependence on the relative momentum forbidden by the interaction, then the rainbow-ladder vector-meson Bethe-Salpeter amplitude takes the form
\begin{equation}
\Gamma_\mu^\rho(P) = \gamma^T_\mu E_\rho(P), \label{rhobsa}
\end{equation}
where $P_\mu \gamma^T_\mu = 0$, $\gamma^T_\mu+\gamma^L_\mu=\gamma_\mu$.  We assume isospin symmetry throughout and hence do not explicitly include the Pauli isospin matrices.\footnote{ Note, too, that we use a Euclidean metric:
$\{\gamma_\mu,\gamma_\nu\} = 2\delta_{\mu\nu}$; $\gamma_\mu^\dagger = \gamma_\mu$; $\gamma_5= \gamma_4\gamma_1\gamma_2\gamma_3$, tr$[\gamma_5\gamma_\mu\gamma_\nu\gamma_\rho\gamma_\sigma]=-4 \epsilon_{\mu\nu\rho\sigma}$; $\sigma_{\mu\nu}=(i/2)[\gamma_\mu,\gamma_\nu]$; $a \cdot b = \mbox{\normalsize $\sum$}_{i=1}^4 a_i b_i$; and $P_\mu$ timelike $\Rightarrow$ $P^2<0$.}

\begin{table}[t]
\caption{Results obtained with $\alpha_{\rm IR}/\pi=0.93$ and (in GeV): $m=0.007$, $\Lambda_{\rm ir} = 0.24\,$, $\Lambda_{\rm uv}=0.905$ \protect\cite{Roberts:2011wy}.  The Bethe-Salpeter amplitudes are canonically normalised; $\kappa_\pi$ is the in-pion condensate \protect\cite{Brodsky:2010xf,Chang:2011mu,Brodsky:2012ku}; and $f_{\pi,\rho}$ are the mesons' leptonic decay constants.  Empirical values are $\kappa_\pi \approx (0.22\,{\rm GeV})^3$ and \protect\cite{Nakamura:2010zzi} $f_\pi=0.092\,$GeV, $f_\rho=0.153\,$GeV.  All dimensioned quantities are listed in GeV.
\label{Table:static}
}
\begin{center}
\begin{tabular*}%{|c|c|c|c|c|c|c|}\hline
{\hsize}
{
c@{\extracolsep{0ptplus1fil}}
c@{\extracolsep{0ptplus1fil}}
c@{\extracolsep{0ptplus1fil}}
|c@{\extracolsep{0ptplus1fil}}
c@{\extracolsep{0ptplus1fil}}
c@{\extracolsep{0ptplus1fil}}
c@{\extracolsep{0ptplus1fil}}
c@{\extracolsep{0ptplus1fil}}
c@{\extracolsep{0ptplus1fil}}
c@{\extracolsep{0ptplus1fil}}}\hline
$E_\pi$ & $F_\pi$ & $E_\rho$ & $M$ & $\kappa_\pi^{1/3}$ & $m_\pi$ & $m_\rho$ & $f_\pi$ & $f_\rho$ \\\hline
%
%0 & 3.568 & 0.459 & 1.520 & 0.358 & 0.241 & 0\,~~~~~ & 0.919 & 0.100 & 0.130\rule{0ex}{2.5ex}\\
%
3.639 & 0.481 & 1.531 & 0.368 & 0.243 & 0.140 & 0.929 & 0.101 & 0.129\\\hline
\end{tabular*}
\end{center}
\end{table}

Values of some meson-related quantities, of relevance herein and computed using the contact-interaction, are reported in Table~\ref{Table:static}.  We quote pion properties in order to provide a broader picture: the pion's Bethe-Salpeter amplitude is
\begin{equation}
\Gamma^\pi(P) = \gamma_5  \bigg[ i E_\pi(P) + \frac{1}{M} \gamma\cdot P F_\pi(P) \bigg].
\end{equation}

\section{$\mathbf{\rho}$-meson Form Factors}
\label{sec:formfactors}
\begin{figure}[t]
\includegraphics[width=0.8\linewidth]{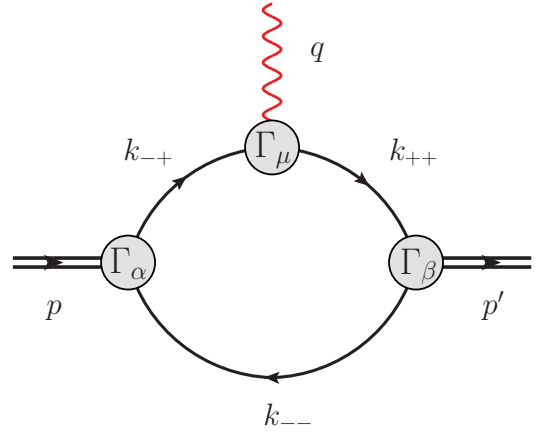}
\caption{Impulse approximation to the $\rho$-$\gamma$ vertex, Eq.\,\protect\eqref{eq:rhoFF}: solid lines -- dressed-quark propagators; and shaded circles, clockwise from top -- Bethe-Salpeter vertex for quark-photon coupling, and Bethe-Salpeter amplitudes for the $\rho^+$-meson.
\label{fig:rhougamma}}
\end{figure}

At this point we can proceed to computation of the form factors.  In order to ensure a symmetry-preserving treatment, one must calculate the vertex in Eq.\,\eqref{rgrvertex} at the same level of approximation as used for the dressed-quark propagator and meson Bethe-Salpeter amplitude; i.e., the generalised impulse approximation:
%\begin{eqnarray}
%\nonumber
%\lefteqn{\Gamma_{\alpha\mu\beta}(p,p^\prime) =
%\int\frac{d^4k}{(2\pi)^4}\,\text{Tr}_{CFD}\bigg\{i\Gamma^{\rho_j}_\beta(k;-p^\prime)
%{\mathpzc S}(k_{++})}\\
%
%&&\times ie\Gamma_\mu(k_{-+},k_{++}){\mathpzc S}(k_{-+})i\Gamma^{\rho_j}_\alpha(k-q/2;p){\mathpzc S}(k_{--})\bigg\},
%\rule{4ex}{0ex}
%\label{GIArho}
%\end{eqnarray}
\begin{eqnarray}
\label{GIArho}
\Gamma_{\alpha\mu\beta}(p,p^\prime) &=&
\Gamma_{\alpha\mu\beta}^u(p,p^\prime) + \Gamma^d_{\alpha\mu\beta}(p,p^\prime)\,,
\end{eqnarray}\vspace*{-8ex}

\begin{eqnarray}
\nonumber
\lefteqn{\Gamma_{\alpha\mu\beta}^f(p,p^\prime) =
2 \int\frac{d^4k}{(2\pi)^4}\,\text{Tr}_{CD}\bigg\{i\Gamma^{\rho_j}_\beta(k;-p^\prime)
S(k_{++})
}\\
%
%&&\times ie\Gamma_\mu^f(k_{-+},k_{++})
&& \times i\Gamma_\mu^f(k_{-+},k_{++})
S(k_{-+})i\Gamma^{\rho_j}_\alpha(k-q/2;p) S(k_{--})\bigg\},
\rule{4ex}{0ex}
\label{eq:rhoFF}
\end{eqnarray}
wherein the trace is over colour and spinor indices and $k_{\alpha\beta}=k + \alpha q/2 + \beta p/2$.
%$e$ is the positron charge,
%and the new element is the dressed-quark--photon vertex.
We illustrate Eq.\,\eqref{eq:rhoFF} in Fig.\,\ref{fig:rhougamma}.

In evaluating Eq.\,\eqref{GIArho} we write:
\begin{equation}
S_f = S + \delta_{\mbox{\st{\scriptsize$CP$}}} \, S_f\,,f=u,d,
\label{Smodified}
\end{equation}
where $S$ is given in Eq.\,\eqref{genS}, with the dressed-mass obtained from Eq.\,\eqref{gapM}, and the broken-$CP$ corrections $\delta_{\mbox{\st{\scriptsize$CP$}}}\, S_f$ are detailed below; and the $\rho$-amplitude
\begin{equation}
\Gamma^{\rho_j}_\alpha = \gamma^T_\alpha  E_\rho(P) +
\Gamma^{\rho_j \mbox{\st{\scriptsize$CP$}}}_\alpha\,,
\label{Gammamodified}
\end{equation}
with $E_\rho(P)$ explained in connection with Eq.\,\eqref{rhobsa} and the broken-$CP$ corrections $\Gamma^{\rho_j \mbox{\st{\scriptsize CP}}}_\alpha$ explained below.  Our computed values for the dressed-quark mass, $M$, and $E_\rho$ are listed in Table~\ref{Table:static}.

The remaining element in Eq.\,\eqref{GIArho} is the dressed-quark--photon vertex.  We are only interested in the $q^2=0$ values of the form factors and hence may use
\begin{eqnarray}
%\Gamma_\mu(p_1,p_2) = {\mathpzc Q} \, \gamma_\mu + i {\mathpzc Q} \, {\mathpzc
e\Gamma_\mu(p_1,p_2) &=& e\tilde{\mathpzc Q} \, \gamma_\mu
%- i {\mathpzc D}\frac{\mathpzc{v}_H}{\Lambda^2} \gamma_5 \sigma_{\mu\nu}(p_2 - p_1)_\nu\,.
+ i \tilde{\mathpzc D} \gamma_5 \sigma_{\mu\nu}(p_2 - p_1)_\nu
\label{qpvEDM}\\
& =: & e\,\mbox{\rm diag}[e_u\Gamma_\mu^u(p_1,p_2),-e_d\Gamma_\mu^d(p_1,p_2)],\;
\end{eqnarray}
where $e$ is the positron charge, $\tilde{\mathpzc Q} = {\rm diag}[e_u=2/3,-e_d=1/3]$
%; $\mathpzc{v}_H=246\,$GeV is the cube-root of the phenomenological Higgs vacuum expectation value;
and
$\tilde{\mathpzc D} = {\rm diag}[d_u, -d_d]$, with $d_f$ the EDM of a current quark with flavour $f$.
%
%I would add that the second CPV term in 23 is only for quark-EDM, for all other EDMs it is given by the second line of DSE2 eq.13, where \bar\Gamma stands for the corresponding expressions given later in the paper.
N.B.\ The second term in Eq.\,\eqref{qpvEDM} describes the explicit current-quark EDM interaction in Eq.\,\eqref{LeffCompute}.  In Sec.\,\ref{sec:formulae} we show that the other terms in Eq.\,\eqref{LeffCompute} generate additional contributions that interfere with this explicit term.

Note that both structures in the vertex, Eq.\,\eqref{qpvEDM}, are in general multiplied by momentum-dependent scalar functions.  Naturally, the vector Ward-Takahashi identity ensures that the coefficient of the $\tilde{\mathpzc Q}\gamma_\mu$ term is ``1'' at $q^2=0$.
In connection with the tensor term, one knows from Ref.\,\cite{Roberts:2011wy} that a tensor vertex is not dressed in the rainbow-ladder treatment of the contact interaction.
However, with a more sophisticated interaction, strong interaction dressing of the $\gamma_5 \sigma_{\mu\nu}$ part of the quark-photon vertex might be significant, given that the dressed-quark-photon vertex certainly possesses a large dressed-quark anomalous magnetic moment term owing to dynamical chiral symmetry breaking \cite{Chang:2010hb}.  At $q^2=0$, this could enhance the strength of the $\tilde {\mathpzc D}$ term by as much as a factor of ten.  If so, then sensitivity to current-quark EDMs is greatly magnified.  It is worth bearing this in mind.

Working with Eq.\,\eqref{rgrvertex}, it is sufficient herein to employ three projection operators:
\begin{subequations}
\begin{eqnarray}
P^1_{\alpha\mu\beta} & = & \mathcal P_{\alpha\sigma}^T(p)P_\mu\mathcal P_{\sigma\beta}^T(p^\prime)\,, \\
\nonumber
P^2_{\alpha\mu\beta} & = & \mathcal P_{\alpha\alpha^\prime}^T(p) \, \mathcal P_{\beta'\beta}^T(p^\prime)\\
&&
\left(
\frac{
\delta_{\mu\beta^\prime}q_{\alpha^\prime}
-
\delta_{\mu\alpha^\prime}q_{\beta^\prime}
}{q^2}
+ \frac{P_\mu\delta_{\alpha^\prime\beta^\prime}}{6p^2}\right) ,
\\
  P^3_{\alpha\mu\beta}&=&\frac{1}{2iq^2}\,\mathcal P_{\alpha\alpha^\prime}^T(p)\varepsilon_{\alpha^\prime\beta^\prime\mu\sigma}q_\sigma \mathcal P_{\beta^\prime\beta}^T(p')\,,
\end{eqnarray}
\end{subequations}
with $p^\prime=p + q$, $P=p + p^\prime$, for then
\begin{subequations}
\begin{eqnarray}
\mathcal E(0) &= & \lim_{q^2\to0}\frac{1}{12m_\rho^2}\,P^1_{\alpha\mu\beta}\Gamma_{\alpha\mu\beta}\,,\\
\mathcal M(0) &= & \lim_{q^2\to0}\frac{1}{4}\,P^2_{\alpha\mu\beta}\Gamma_{\alpha\mu\beta}\,, \\
\mathcal D(0) &=& %\lim_{q^2\to0}\frac{1}{2m_\rho}\,P^3_{\alpha\mu\beta}\Gamma_{\alpha\mu\beta}\,.
\lim_{q^2\to0} P^3_{\alpha\mu\beta}\Gamma_{\alpha\mu\beta} \,,
\end{eqnarray}
\end{subequations}
and $\mu_\rho = {\cal M}(0)\, e/[2 m_\rho]$, $d_\rho = {\cal D}(0) \, e/[2 m_\rho]$.
So long as a global-symmetry-preserving regularisation scheme is implemented, $\mathcal E(0)=1$; the value of $\mathcal M(0)$ is then a prediction, which can both be compared with that produced by other authors and serve as a benchmark for our prediction of $\mathcal D(0)$.

At this point one has sufficient information to calculate the $\rho$-meson's magnetic moment.  We simplify the denominator in Eq.\,\eqref{GIArho} via a Feynman parametrisation:
\begin{eqnarray}
\nonumber
\lefteqn{
\left(k_{++}^2 + M^2\right)^{-1}\left(k_{-+}^2 + M^2\right)^{-1}\left(k_{--}^2 + M^2\right)^{-1}}\\
%
%\nonumber
%&=& 2\int_0^1\int_0^{1-x} \!\! dx\, dy \bigg[ x k_{+-}^2 + y k_{-+}^2  \\
%
%\nonumber
%&& \rule{10em}{0ex} + (1 - x -y) k_{++}^2 + M^2 \bigg]^{-3} \\
%
\nonumber
&=& 2\int_0^1\int_0^{1-x}\!\! dx\,dy \bigg[k^2 + M^2 \\
\nonumber
&& \quad + \frac{1}{4} \left[p^2 - 2\,(1 - 2x - 2y)\,p\cdot q  + q^2 \right] \\
&& \quad - (1 - 2y)\,q\cdot k + (1 - 2x)\,p\cdot k \bigg]^{-3} .
\end{eqnarray}
This appears as part of an expression that is integrated over four-dimensional $k$-space.  The expression is simplified by a shift in integration variables, which exposes a  denominator of the form $1/[k^2 + \tilde M^2]^3$, with
\begin{equation}
\tilde M^2=M^2 + x(x - 1)\,m_\rho^2 + y(1 - x - y)\,Q^2\,.
\end{equation}

One thereby arrives at a compound expression that involves one-dimensional integrals of the form in Eq.\,\eqref{gapM}, which we regularise via Eq.\,\eqref{RegC} and generalisations thereof; viz.,
\begin{subequations}
\begin{eqnarray}
\int ds\,\frac{s}{[s + \omega]^2}&=&
-\frac{d}{d\omega}\,\mathcal C^\text{iu}(\omega)
=: \overline{\mathcal C}_1^\text{iu}(\omega)\,, \\
\int ds\,\frac{s}{[s + \omega]^3}&=&
\frac{1}{2}\frac{d^2}{d\omega^2}\,\mathcal C^\text{iu}(\omega)
=: \overline{\mathcal C}_2^\text{iu}(\omega)\,, \\
\int ds\,\frac{s^2}{[s + \omega]^3}&=&
\overline{\mathcal C}_1^\text{iu}(\omega) -
\omega\overline{\mathcal C}_2^\text{iu}(\omega)\,,
%
%\int_R ds\,\frac{s^3}{\left(s + \tilde M^2\right)^3}&=&
%\mathcal C^\text{iu}(\omega)\Big|_{\omega=\tilde M^2} - 2\,\mathcal C_1^\text{iu}(\omega)\Big|_{\omega=\tilde M^2} + \mathcal C_2^\text{iu}(\omega)\Big|_{\omega=\tilde M^2}\,.
\end{eqnarray}
\end{subequations}
etc.  Details for this component of our computation may be found in Ref.\,\cite{Roberts:2011wy} and pursuing it to completion one obtains the magnetic moment listed in Table~\ref{Table:magmom}.

\begin{figure}[t]
\centerline{%
\includegraphics[clip,width=0.4\textwidth]{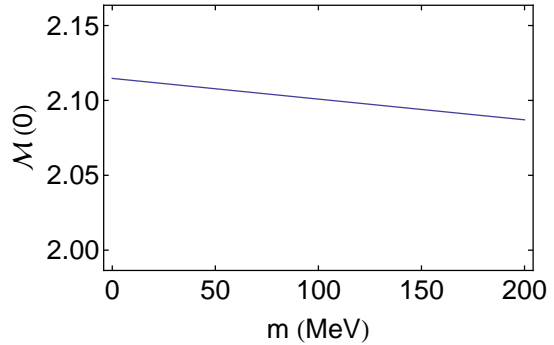}}
\caption{\label{fig:MDM} Evolution of $\rho$-meson magnetic moment with current-quark mass. $m=170\,$MeV corresponds to the mass of the $s$-quark in our treatment of the contact interaction \protect\cite{Chen:2012qr}, so the difference between $\mathpzc M_\rho(0)$ and $\mathpzc M_\phi(0)$ is just 1\%.}
\end{figure}

\begin{table}[b]
\caption{Magnetic moment of the $\rho$-meson calculated using our framework; and a comparison with other computations.
\underline{Legend}: \emph{RL RGI-improved}, treatment of a renormalisation-group-improved one-gluon exchange kernel in rainbow-ladder truncation;
\emph{EF parametrisation}, entire function parametrisation of solutions to the gap and Bethe-Salpeter equations;
and \emph{LF CQM}, light-front constituent-quark model.
The results are listed in units of $e/[2m_\rho]$.
\label{Table:magmom}
}
\begin{center}
\begin{tabular*}%{|c|c|c|c|c|c|c|}\hline
{\hsize}
{
l@{\extracolsep{0ptplus1fil}}
l@{\extracolsep{0ptplus1fil}}}\hline
This work and Ref.\,\cite{Roberts:2011wy} & 2.11\\
DSE: RL RGI-improved  \cite{Bhagwat:2006pu} & 2.01 \\
DSE: EF parametrisation \cite{Hawes:1998bz}   & 2.69 \\
LF CQM \cite{deMelo:1997hh} & 2.14 \\
LF CQM \cite{Choi:2004ww} & 1.92 \\
Sum Rules \cite{Samsonov:2003hs} & $1.8 \pm 0.3$\\
point particle & 2 \\\hline
\end{tabular*}
\end{center}
\end{table}

We depict the evolution of $\mathpzc M(0)$ with current-quark mass in Fig.\,\ref{fig:MDM}: $\mathpzc M(0)$ is almost independent of $m$.  This outcome matches that obtained in Ref.\,\cite{Bhagwat:2006pu} using a renormalisation-group-improved one-gluon exchange kernel and hence a momentum-dependent dressed-quark mass-function of the type possessed by QCD \cite{Bhagwat:2003vw,Bowman:2005vx,Bhagwat:2006tu,Bhagwat:2007vx}.  The behaviour in Fig.\,\ref{fig:MDM} will serve to benchmark that of the $\rho$-meson's EDM.

\section{$\mathbf{\rho}$-meson EDM: Formulae}
\label{sec:formulae}
%---Effective operator -- produces three terms
%---EDM and CEDM operators -- 3 terms
%---theta => 1 term
\label{rhoEDM}
We now turn to computation of the effect of the interaction terms in Eq.\,\eqref{LeffCompute} on the $\rho$-meson.
There are three types of contribution, which arise separately through modification of:
(1) the quark-photon vertex, Eq.\,\eqref{qpvEDM};
(2) the $\rho$-meson Bethe-Salpeter amplitude, Eq.\,\eqref{Gammamodified}; and
(3) the dressed-quark propagator, Eq.\,\eqref{Smodified}.
%
%Each term in Eq.\,\eqref{LeffCompute} can effect all such alterations, so we analyse their contributions separately.

\subsection{Four-fermion interaction}
We begin with the dimension-six operator, which can be written explicitly as
\begin{eqnarray}
\nonumber
\mathcal L_6 & = & i \frac{{\cal K}}{2{\Lambda^2}}
    \left[
  \bar u^ad^a\bar d^b\gamma_5u^b + \bar u^a\gamma_5d^a\bar d^bu^b \right.\\
&&   \quad \left. - \bar d^ad^a\bar u^b\gamma_5u^b - \bar d^a\gamma_5d^a\bar u^bu^b\right], \label{L6def}
\end{eqnarray}
with summation over the repeated colour indices.  This operator generates all three types of modification.

\begin{figure}[t]
\centerline{
\includegraphics[width=0.38\linewidth]{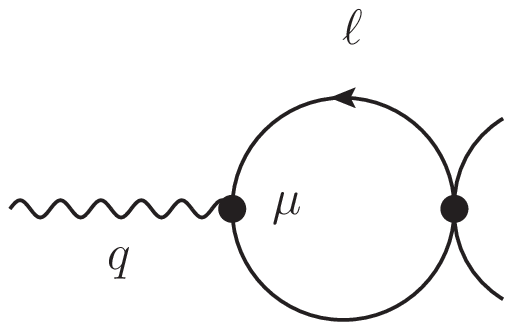}}
\vspace*{3ex}

\centerline{
\includegraphics[width=0.38\linewidth]{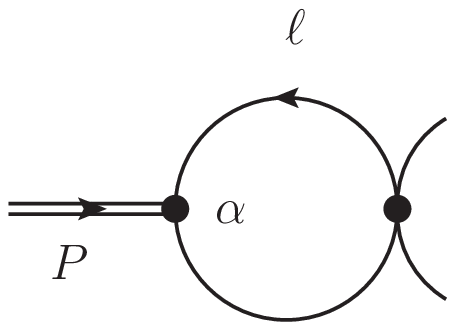}}
\vspace*{3ex}

\centerline{
\includegraphics[width=0.38\linewidth]{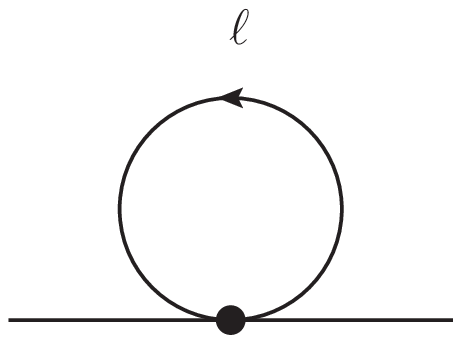}}

\caption{\emph{Top} -- Correction to the quark-photon vertex generated by the four-fermion operator in Eq.\,\protect\eqref{L6def}.  The unmodified quark-photon vertex is the left dot, whereas the right dot locates insertion of $\mathpzc L_6$.  If the internal line represents a circulating $d$-quark then, owing to the $\mathpzc L_6$ insertion, the external lines are $u$-quarks, and vice versa.\label{EBS2}
\emph{Middle} -- Analogous correction to the $\rho$-meson Bethe-Salpeter amplitude.  The unmodified amplitude is the left dot, whereas the right dot locates insertion of $\mathpzc L_6$.  The lower internal line is an incoming $d$-quark and the upper external line is an outgoing $u$-quark.
\emph{Bottom} -- $\mathpzc L_6$-correction to the dressed-quark propagator, with the dot locating the operator insertion.  If the outer line is a $u$-quark, then the internal line is a $d$-quark; and vice versa.
%\label{EBS}%
}
\end{figure}

\subsubsection{$\mathcal L_6$ -- quark-photon vertex}
\label{sss:L6QPV}
This contribution is depicted in the top panel of Fig.\,\ref{EBS2}.  Consider first the case of $d$-quarks circulating in the loop, then straightforward but careful analysis of the induced Wick contractions produces the following result:
\begin{subequations}
\begin{eqnarray}
\label{QPBS1}
\Gamma_\mu^{\gamma_{\mathcal L_6^d}} &=& -i\frac{\cal K}{\Lambda^2}
\frac{e_d}{e_u}
\int\frac{d^4\ell}{(2\pi)^4}\big[ \mathcal{I}_\mu^{12} + N_c \mathpzc I_\mu^3 \big]\,,\\
\mathpzc I_\mu^{12} & = &  -P_RS(\ell + q)\gamma_\mu S(\ell)P_R \nonumber \\
&& \quad + P_LS(\ell + q)\gamma_\mu S(\ell)P_L\,, \label{calI12} \\
\mathpzc I_\mu^3&=& P_L\,\text{tr}\{S(\ell + q)\gamma_\mu S(\ell)P_L\} \nonumber \\
&& \quad - P_R\,\text{tr}\{S(\ell + q)\gamma_\mu S(\ell)P_R\} \,,\label{calI3}
\end{eqnarray}
\end{subequations}
where $P_{R,L}=(1/2)(1 \pm \gamma_5)$.  These right- and left-handed projection operators satisfy $P_R+P_L = I_{\rm D}$.

Further simplification of the integrand reveals
\begin{subequations}
\begin{eqnarray}
\mathpzc I_\mu^{12} & = & \mathpzc I_\mu^{1} + \mathpzc I_\mu^{2} \nonumber \\
&=& \frac{i\gamma\cdot q}{(\ell + q)^2 + M^2}\gamma_\mu\frac{M}{\ell^2 + M^2}\gamma_5 \\
&& \quad + 2i\frac{\ell_\mu}{(\ell + q)^2 + M^2}\frac{M}{\ell^2 + M^2}\gamma_5 \,, \\
\mathpzc I_\mu^3 & = & \frac{2i(2\ell_\mu + q_\mu)}{(\ell + q)^2 + M^2}\frac{M}{\ell^2 + M^2}\gamma_5\,,
\end{eqnarray}
\end{subequations}
so that one may subsequently obtain
\begin{subequations}
\begin{eqnarray}
\int\frac{d^4\ell}{(2\pi)^4} \mathpzc I_\mu^{1}
&=& (q_\mu + i\sigma_{\mu\nu}q_\nu)\gamma_5 \nonumber \\
&& \times  \frac{i M}{16\pi^2}\int_0^1dx\,\overline {\mathpzc C}_1^{\rm iu}(\omega_q)
%x(1 -  x)q^2 + M^2)
\,, \\
\int\frac{d^4\ell}{(2\pi)^4} \mathpzc I_\mu^{2}
&=& -q_\mu\gamma_5 \frac{i M}{8\pi^2}\int_0^1dx\,x\,\overline{\mathpzc C}_1^{\rm iu}(\omega_q)\,,\\
\int\frac{d^4\ell}{(2\pi)^4} \mathpzc I_\mu^{3}
&= &q_\mu\gamma_5
\frac{i M}{8\pi^2}\int_0^1dx\,(1-2x)\,\overline{\mathpzc C}_1^{\rm iu}(\omega_q)\,,
\rule{2ex}{0ex}
\end{eqnarray}
\end{subequations}
where $\omega_q=x(1 -  x)q^2 + M^2$.
Combining the terms, Eq.\,\eqref{QPBS1} becomes
\begin{eqnarray}
\nonumber \Gamma_\mu^{\gamma_{\mathcal L_6^d}} &=&
\frac{\cal K}{\Lambda^2}
\frac{e_d}{e_u}\frac{M}{16\pi^2} \int_0^1 \! dx \,
\overline{\mathpzc C}_1^{\rm iu}(\omega_q)\\
&& \times [(1+2N_c) (1-2 x) q_\mu + i\sigma_{\mu\nu}q_\nu ] \gamma_5 \\
&\stackrel{q^2=0}{=} &
\frac{\cal K}{\Lambda^2}
\frac{e_d}{e_u}\frac{M}{16\pi^2} \overline{\mathpzc C}_1^{\rm iu}(M^2)
i\sigma_{\mu\nu}q_\nu  \gamma_5\,.
\end{eqnarray}

In the other case, with a $u$-quark circulating in the loop, one obtains
\begin{eqnarray}
\Gamma_\mu^{\gamma_{\mathcal L_6^u}}
&\stackrel{q^2=0}{=} &
\frac{\cal K}{\Lambda^2}
\frac{e_u}{e_d}\frac{M}{16\pi^2} \overline{\mathpzc C}_1^{\rm iu}(M^2)
i\sigma_{\mu\nu}q_\nu  \gamma_5\,.
\end{eqnarray}

Plainly, the net correction to the quark-photon vertex can now be cast in the form of the second term in Eq.\,\eqref{qpvEDM} and hence is readily expressed in ${\cal D}(0)$.

\subsubsection{$\mathcal L_6$ -- Bethe-Salpeter amplitude}
\label{sss:L6BSA}
This correction is depicted in the middle panel of Fig.\,\ref{EBS2}.  Each of the four terms in Eq.\,\eqref{L6def} generates a distinct contribution.  That from the first and second are:
\begin{subequations}
\begin{eqnarray}
\Gamma_\alpha^{\rho \mathcal L_6^1}
&=& -i\frac{\cal K}{\Lambda^2} N_c E_\rho\,P_R\, \nonumber \\
&& \quad \times \text{tr}\int\frac{d^4\ell}{(2\pi)^4}S(\ell)P_RS(\ell + P)\gamma_\alpha^T\,, \\
\Gamma_\alpha^{\rho \mathcal L_6^2}
&=& -i\frac{\cal K}{\Lambda^2} E_\rho\, P_R\, \nonumber \\
&& \quad \times \int\frac{d^4\ell}{(2\pi)^4}S(\ell + P)\gamma_\alpha^T S(\ell)P_R\,.
\end{eqnarray}
\end{subequations}
The third and fourth terms are identical, up to sign-change and the replacement $P_R\to P_L$; and hence
\begin{eqnarray}
\Gamma_\alpha^{\rho \mathcal L_6} & = & i\frac{\cal K}{\Lambda^2} E_\rho
\int\frac{d^4\ell}{(2\pi)^4}\big[ \mathcal{I}_\alpha^{12{\rm T}} + N_c \mathpzc I_\alpha^{3{\rm T}} \big],
\end{eqnarray}
where the superscript ``T'' indicates that $\gamma_\alpha^{\rm T}$ is here used in the expressions for $\mathcal{I}^{12}$, $\mathcal{I}^{3}$.

Now, using the formulae of Sec.\,\ref{sss:L6QPV}, one arrives at
\begin{equation}
\Gamma_\alpha^{\rho \mathcal L_6} = -i \frac{\cal K}{\Lambda^2}
\frac{M E_\rho}{16\pi^2} \gamma_5\sigma_{\alpha\nu}P_\nu
\int_0^1\!dx\, \overline{\mathpzc C}_1^{\rm iu}(\omega_P)\,,
\end{equation}
where $\omega_P=x(1-x)P^2+M^2$, $P^2=-m_\rho^2$.  This is one of the additive corrections to the Bethe-Salpeter amplitude anticipated in Eq.\,\eqref{Gammamodified}.

\subsubsection{$\mathcal L_6$ -- quark propagator}
\label{sss:L6S}
The final modification arising from the dimension-six operator is that depicted in the bottom panel of Fig.\,\ref{EBS2}.  So long as the correction is small, it modifies the dressed-quark propagator as follows:
\begin{equation}
S(k) \to S(k)+ \delta_{\mathcal L_6}S(k) = S(k) + S(k) i\Gamma^{S \mathcal L_6} S(k)\,,
\end{equation}
where, once again, each of the four terms in Eq.\,\protect\eqref{L6def} contributes.  Their sum is
\begin{eqnarray}
\nonumber
\lefteqn{\Gamma^{S \mathcal L_6} =
\frac{\cal K}{\Lambda^2} \int\frac{d^4\ell}{(2\pi)^4}
\big[
P_R S(\ell)P_R - P_LS(\ell)P_L }\\
&& +
N_cP_R\,\text{tr}\{S(\ell)P_R\}
- N_cP_L\,\text{tr}\{S(\ell)P_L\}\big]\,.
\end{eqnarray}

Now
\begin{eqnarray}
\nonumber\lefteqn{
P_R S(\ell)P_R - P_LS(\ell)P_L  = \frac{M}{\ell^2 + M^2} \gamma_5}\\
& = & \frac{1}{2} \big[ P_R\,\text{tr}\{S(\ell)P_R\}
- P_L\,\text{tr}\{S(\ell)P_L\} \big],
\end{eqnarray}
so that with little additional algebra one arrives at
\begin{equation}
\delta_{\mathcal L_6}S(k) = \frac{i}{k^2+M^2}(1+2N_c)\frac{\cal K}{\Lambda^2}
\frac{M}{16\pi^2} \mathpzc C^{\rm iu}(M^2)\gamma_5\,.
\label{deltaSL6}
\end{equation}

\subsection{Quark chromo-EDM}
The term in the middle line of Eq.\,\eqref{LeffCompute} also generates all three types of modification described in the opening lines of this Section.  Notably, owing to dynamical chiral symmetry breaking, the dressed-quark-gluon coupling possesses a chromomagnetic moment term that, at infrared momenta, is two orders-of-magnitude larger than the perturbative estimate \cite{Chang:2010hb}.  One may reasonably expect similar strong-interaction dressing of a light-quark's chromo-EDM interaction with a gluon, in which case sensitivity to the current-quark's chromo-EDM is very much enhanced.

\subsubsection{$\mathpzc L_{CEDM}$ -- quark-photon vertex}
\label{sss:LCEDMQPV}
\begin{figure}[t]
\centerline{
\includegraphics[width=0.99\linewidth]{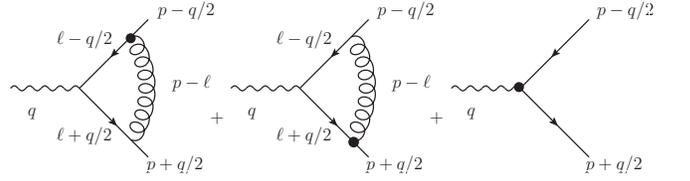}}
\caption{Correction to the quark-photon vertex generated by the quark chromo-EDM operator in Eq.\,\protect\eqref{LeffCompute}: the incoming and outgoing quark lines have the same flavour, $f$.  The dot in the left two diagrams locates insertion of $\mathpzc L_{CEDM}$, whilst that in the rightmost diagram indicates the second term in Eq.\,\protect\eqref{qpvEDM}; i.e., the explicit quark EDM. \label{EBS5}}
\end{figure}

This contribution is depicted in Fig.\,\ref{EBS5}.  After a lengthy analysis, in which we represent the exchanged gluon via Eq.\,\eqref{njlgluon}, the sum of the two leftmost diagrams produces
%\begin{subequations}
%\begin{eqnarray}
%\Gamma^{\gamma(g)}_\mu&=&\Gamma^{\gamma (g) 1}_\mu+\Gamma^{\gamma 2(g) }_\mu \,, \\
%
%\Gamma^{\gamma (g) 1}_\mu&=&
% \frac{i}{6\pi}\frac{\tilde d_f \alpha_{\rm IR}}{m_G^2}\int_0^1dx\left[\mathpzc C^{\rm iu}(\omega_q) - {\mathpzc C}_1^{\rm iu}(\omega_q)\right] \nonumber\\
%
%\nonumber   &&\times \bigg\{2q_\alpha\,\sigma_{\mu\alpha}\gamma_5 - 6i\big[3(x - 1/2)q_\mu - p_\mu\big]\,\gamma_5\bigg\}\,, \\
%&&\\
%
%\Gamma^{(g)\gamma 2}_\mu&=& \frac{1}{3\pi}\frac{\tilde d_f\alpha_{\rm IR}}{m_G^2}
%\int_0^1dx\,\overline{\mathpzc C}_1^{\rm iu}(\omega_q)
%\bigg\{6 \big[\omega_q - 2 M^2\big]\,p_\mu\gamma_5 \nonumber\\
%  && -6\big[(x - 1/2)\omega_q + 2x(1 - x)q\cdot p\big]\,q_\mu\gamma_5 \nonumber \\
%  &&  + M\big[((x - 1/2)q + p)\cdot\gamma\big]q_\alpha\sigma_{\alpha\mu}\gamma_5 \nonumber\\
%  &&  + Mq_\alpha\sigma_{\alpha\mu}\gamma_5\big[((x - 1/2)q + p)\cdot\gamma\big]\bigg\}\,,
%\end{eqnarray}
%\end{subequations}
\begin{eqnarray}
\Gamma^{\gamma(g)}_\mu &=&
 \frac{1}{6i\pi}\frac{\tilde d_f \alpha_{\rm IR}}{m_G^2}\int_0^1dx\left[\mathpzc C^{\rm iu}(\omega_q) - {\mathpzc C}_1^{\rm iu}(\omega_q)\right] \nonumber\\
\nonumber   &&\times \bigg\{2q_\alpha\,\sigma_{\mu\alpha}\gamma_5 - 6i\big[3(x - 1/2)q_\mu - p_\mu\big]\,\gamma_5\bigg\} \nonumber \\
&-& \frac{1}{3\pi}\frac{\tilde d_f\alpha_{\rm IR}}{m_G^2}
\int_0^1dx\,\overline{\mathpzc C}_1^{\rm iu}(\omega_q)
\bigg\{6 \big[\omega_q - 2 M^2\big]\,p_\mu\gamma_5 \nonumber\\
  && -6\big[(x - 1/2)\omega_q + 2x(1 - x)q\cdot p\big]\,q_\mu\gamma_5 \nonumber \\
  &&  + M\big[((x - 1/2)q + p)\cdot\gamma\big]q_\alpha\sigma_{\alpha\mu}\gamma_5 \nonumber\\
  &&  + Mq_\alpha\sigma_{\alpha\mu}\gamma_5\big[((x - 1/2)q + p)\cdot\gamma\big]\bigg\}\,,
\end{eqnarray}
where, again, $\tilde d_f$ is the chromo-EDM of a quark with flavour $f$.

As we are interested solely in the EDM, we may consider $q^2=0$, at which value the result simplifies greatly:
%\begin{subequations}
%\begin{eqnarray}
%\Gamma^{\gamma (g) 1}_\mu &=&
% \frac{1}{3\pi}\frac{\tilde d_f\alpha_{\rm IR}}{m_G^2}
% \big[\mathpzc C^{\rm iu}(M^2) - {\mathpzc C}_1^{\rm iu}(M^2)\big]  \nonumber\\
%&& \quad \times \big[i \gamma_5\sigma_{\mu\alpha}q_\alpha - 3 p_\mu\,\gamma_5\big] \,,\\
%
%\Gamma^{\gamma (g) 2}_\mu&=& -\frac{1}{3\pi}\frac{\tilde d_f\alpha_{\rm IR}}{m_G^2}\,
%\overline{\mathpzc C}_1^{\rm iu}(M^2)\,
%\bigg[M \{\gamma\cdot p, \gamma_5 \sigma_{\mu\alpha}\} q_\alpha
%\nonumber\\
%
%&& \quad + 2 p\cdot q q_\mu \gamma_5 + 6 M^2 p_\mu \gamma_5 \bigg].
%\end{eqnarray}
%\end{subequations}
\begin{eqnarray}
\Gamma^{\gamma (g)}_\mu &=&
 \frac{1}{3i\pi}\frac{\tilde d_f\alpha_{\rm IR}}{m_G^2}
 \big[\mathpzc C^{\rm iu}(M^2) - {\mathpzc C}_1^{\rm iu}(M^2)\big]  \nonumber\\
&& \quad \times \big[\gamma_5\sigma_{\mu\alpha}q_\alpha + 3i p_\mu\,\gamma_5\big] \nonumber\\
&+& \frac{1}{3\pi}\frac{\tilde d_f\alpha_{\rm IR}}{m_G^2}\,
\overline{\mathpzc C}_1^{\rm iu}(M^2)\,
\bigg[M \{\gamma\cdot p, \gamma_5 \sigma_{\mu\alpha}\} q_\alpha
\nonumber\\
&& \quad + 2 p\cdot q q_\mu \gamma_5 + 6 M^2 p_\mu \gamma_5 \bigg].
\end{eqnarray}

Plainly, the net correction to the quark-photon vertex from these two diagrams can now be cast in the form of the second term in Eq.\,\eqref{qpvEDM}, which, in fact, is precisely the rightmost diagram in Fig.\,\ref{EBS5} because $q=p_2-p_1$.

\subsubsection{$\mathpzc L_{CEDM}$ -- Bethe-Salpeter amplitude}
\label{sss:LCEDMBSA}
\begin{figure}[t]
\centerline{
\includegraphics[width=0.95\linewidth]{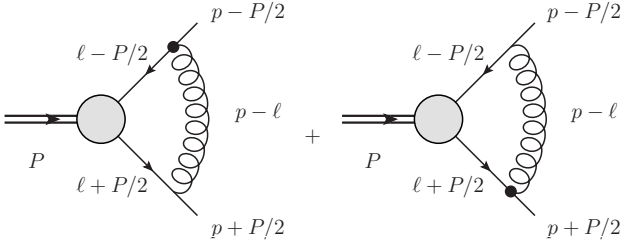}}
\caption{Correction to the $\rho$-meson Bethe-Salpeter amplitude generated by the quark chromo-EDM operator in Eq.\,\protect\eqref{LeffCompute}: the incoming line is a $d$-quark and the outgoing line is a $u$-quark.  In each case the dot locates insertion of $\mathpzc L_{CEDM}$. \label{EBS4}}
\end{figure}

This correction is expressed in Fig.\,\ref{EBS4}.  Owing to similarity between the $\mathpzc L_{\rm eff}$-uncorrected $\rho$-meson amplitude and quark-photon vertex, the results can be read from those in Sec.\,\ref{sss:LCEDMQPV}; viz., with $\tilde d_\pm = \tilde d_u \pm \tilde d_d$,
\begin{eqnarray}
\lefteqn{\Gamma^{\rho(g)}_{\alpha} =
\frac{1}{6i\pi}\frac{\alpha_{\rm IR}}{m_G^2}E_\rho\int_0^1dx
\left[\mathpzc C^{\rm iu}(\omega_P) - \mathpzc C_1^{\rm iu}(\omega_P)\right] }\nonumber\\
&&\quad \times \bigg\{
\big[(\tilde d_+ - 3 (x-1/2) \tilde d_- )P_\beta
- \tilde d_-p_\beta\big]\,\sigma_{\mu\beta}\gamma_5\mathcal P_{\mu\alpha}^T \nonumber\\
  &&\quad+ 3i\tilde d_+p_\mu\gamma_5\mathcal P_{\mu\alpha}^T
  - 3\tilde d_-M\gamma_\mu\gamma_5\mathcal P_{\mu\alpha}^T\bigg\}\nonumber\\
  &-&\frac{1}{3\pi}\frac{\alpha_{\rm IR}}{m_G^2}E_\rho\int_0^1dx\,\bar {\mathpzc C}_1^{\rm iu}(\omega_P)\nonumber\\
&& \times \bigg\{
  3\tilde d_+\big[\omega_P-2M^2\big]\,p_\mu\gamma_5\mathcal P_{\mu\alpha}^T  \nonumber\\
&&\quad - \tilde d_-\big([\omega_P - 2 M^2][(x - 1/2)P_\beta + p_\beta]\big)\,i\gamma_5\sigma_{\beta\mu}\mathcal P_{\mu\alpha}^T \nonumber\\
&&\quad + M\tilde d_d\big[((x - 1/2)P + p)\cdot\gamma\big]P_\beta\sigma_{\beta\mu}\gamma_5\mathcal P_{\mu\alpha}^T \nonumber\\
&&\quad + M\tilde d_u P_\beta\sigma_{\beta\mu}\gamma_5\mathcal P_{\mu\alpha}^T\big[((x - 1/2)P + p)\cdot\gamma\big]
  \bigg\}. \label{neededQCEDM}
\end{eqnarray}

In computing the vertex in Eq.\,\eqref{GIArho} one must employ Fig.\,\ref{EBS4} and also its charge conjugate, the form of which is obtained from Eq.\,\eqref{neededQCEDM} via the interchange $\tilde d_u \leftrightarrow \tilde d_d$, and $p \to -p$, $P\to -P$.

\subsubsection{$\mathpzc L_{CEDM}$ -- quark propagator}
\label{sss:LCEDMS}
\begin{figure}[t]
\centerline{
\includegraphics[width=0.95\linewidth]{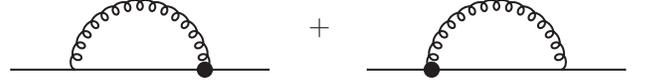}}
\caption{Correction to the dressed-quark propagator generated by the quark chromo-EDM operator in Eq.\,\protect\eqref{LeffCompute}.  In each image the dot locates insertion of $\mathpzc L_{CEDM}$. \label{EBS6}}
\end{figure}

The last modification generated by the chromo-EDM term in Eq.\,\eqref{LeffCompute} is that to the quark propagator, Fig.\,\ref{EBS6}.  The self-energy insertion is readily evaluated:
\begin{equation}
\Gamma^{S(g)} = \tilde d_f \, \frac{8}{\pi} \, \frac{\alpha_{\rm IR}}{m_G^2}\, \mathpzc D^{\rm iu} (M^2) \gamma_5\,,
\end{equation}
where
\begin{equation}
\mathpzc D^{\rm iu}(\omega) = \int ds\frac{s^2}{s+\omega} \to \int_{\tau_{\rm uv}^2}^{\tau_{\rm ir}^2} d\tau\,\frac{2}{\tau^3}\,\exp(-\tau\omega),
\end{equation}
so that, with $f=u,d$,
\begin{equation}
\delta_{(g)}S_f(k) = \frac{i}{k^2+M^2} \, \tilde d_f \, \frac{8}{\pi} \, \frac{\alpha_{\rm IR}}{m_G^2}\, \mathpzc D^{\rm iu}(M^2) \gamma_5\,.
\label{deltaSg}
\end{equation}

\subsection{$\mathbf\theta$-term}
%Owing to a connection between the Higgs mechanism for generating current-quark masses in the SM, \emph{CP} violation in the weak interaction and the $V-A$ character of that theory, the effect of the $\theta$-term can completely be expressed through a $U_A(1)$ rotation of the current-quark mass-matrix.
Owing to a connection between the Higgs mechanism for generating current-quark masses in the SM and \emph{CP} violation in the weak interaction, the effect of the $\theta$-term can completely be expressed through a $U_A(1)$ rotation of the current-quark mass-matrix.
We consider the $s$-quark to be massive and $m_u=m_d$, in which case the effect of the first term in Eq.\,\eqref{LeffCompute} is expressed simply in a modification of the dressed-quark propagator:
\begin{eqnarray}
S(k) &\to &\frac{1}{i \gamma\cdot k + M + \frac{i}{2} m \, \bar\theta \, \gamma_5}\\
& \stackrel{m\bar\theta\,{\rm small}} \approx & S(k) - \frac{1}{k^2+M^2} \frac{i}{2} m \, \bar\theta \, \gamma_5 \,.
\label{deltaStheta}
\end{eqnarray}

\subsubsection{Dressed-quark anomalous chromomagnetic moment}
\label{sec:ACM}
In our global-symmetry-preserving rainbow-ladder treatment of the contact interaction, the general form of the $\rho$-meson's Bethe-Salpeter amplitude is given in Eq.\,\eqref{rhobsa}.  The absence of a term $\sigma_{\mu\nu}P_\nu F_\rho(P)$ is an artefact of the rainbow-ladder truncation: even using Eq.\,\eqref{njlgluon}, a Bethe-Salpeter amplitude with $F_\rho(P)\neq 0$ is obtained in any symmetry-preserving truncation that goes beyond this leading order \cite{Bender:1996bb}.  One material consequence of this omission is complete cancellation of all terms at leading-order in $\bar\theta$, so that the $\theta$-term's contribution to the $\rho$-meson's EDM is anomalously suppressed in rainbow-ladder truncation.  This defect may be ameliorated by acknowledging that the dressed-quark-gluon vertex possesses an anomalous chromomagnetic moment coupling which is enhanced by dynamical chiral symmetry breaking \cite{Chang:2010hb}.  We therefore include an effect generated by
\begin{equation}
\label{vtxACM}
\Gamma_\mu^{\rm acm}(p_i,p_f) = \frac{\mu^{\rm acm}}{2 M} \sigma_{\mu\nu} (p_f-p_i)_\nu\,,
\end{equation}
where \cite{Chang:2011ei} $\mu^{\rm acm} \sim (-1/4)$.

In order to explicate the effect we find it convenient to first express collectively the corrections to the dressed-quark propagator computed above; viz., from Eqs.\,\eqref{deltaSL6}, \eqref{deltaSg}, \eqref{deltaStheta},
\begin{eqnarray}
S(k) & \to & S(k) - i \gamma_5 \frac{\lambda}{k^2+M^2}\,,\\
\lambda_{{\cal L}_6}
& = & - (1+2 N_c) \frac{\cal K}{\Lambda^2} \frac{M}{16 \pi^2} {\cal C}^{\rm iu}(M^2)\,,
\label{lambda1}\\
\lambda_{(g)}
& = & -\tilde d_f \, \frac{8}{\pi} \, \frac{\alpha_{\rm IR}}{m_G^2}\, \mathpzc D^{\rm iu}(M^2)\,,
\label{lambda2}\\
\lambda_{\bar \theta} & = & \frac{1}{2} m \bar \theta\,.
\label{lambda3}
\end{eqnarray}

Our corrections are now obtained via the diagrams in Fig.\,\ref{EBS4}, except that here the dots represent Eq.\,\eqref{vtxACM}, and one simultaneously adds the correction to one and then the other propagator.  In this way, careful but straightforward computation yields
%\begin{eqnarray}
%\Gamma_\mu^{\lambda, {\rm acm}} & = &
%\Gamma_\mu^{\lambda, {\rm acm}_1}+\Gamma_\mu^{\lambda {\rm acm}_2}\,,\\
%
%\Gamma_\mu^{\lambda, {\rm acm}_1}
%
%&=& \frac{i\alpha_{\rm IR}}{2\pi m_G^2}\frac{\lambda^1\mu_2^{\rm acm} - \lambda^2\mu_1^{\rm acm}}{2 M}
%\nonumber\\
%
%&& \times
%\int_0^1dx \left[C^{\rm iu}(\omega_P) - C_1^{\rm iu}(\omega_P)\right]\gamma_\mu\gamma_5 \,,\\
%%
%\Gamma_\mu^{\lambda, {\rm acm}_2} &=&
%\frac{i \alpha_{\rm IR}}{6\pi m_G^2}\frac{1}{M}
%\int_0^1dx\,\bar C_1^{\rm iu}(\omega_P) \nonumber \\
%
%&\times &
%\bigg\{3\mu_-^{\rm acm}\gamma\cdot(p + (x - 1/2)P)\nonumber \\
%&& \quad \times [(1 - x)\lambda^1 - x\lambda^2]P_\mu  \nonumber \\
%
%&& + i\big[(1 - x)\lambda^1
%+ x\lambda^2\big)] \big[\mu_1^{\rm acm}\gamma_\nu P_\alpha\sigma_{\alpha\mu} \nonumber\\
%
%&& \quad - \mu_2^{\rm acm}P_\alpha\sigma_{\alpha\mu}\gamma_\nu\big](p + (x - 1/2)P)_\nu \nonumber\\
%
%&&- \lambda^-M \big[\mu_+^{\rm acm}(p + (x - 1/2)P)_\beta\sigma_{\mu\beta} \nonumber \\
%
%&& \quad + 3i\mu_-^{\rm acm}(p + (x - 1/2)P)_\mu\big]\bigg\}\,\gamma_5\,, \label{correctionmuacm}
%\end{eqnarray}
\begin{eqnarray}
\Gamma_\mu^{\lambda, {\rm acm}} & = &
\frac{\alpha_{\rm IR}}{2i\pi m_G^2}\frac{\lambda^1\mu_2^{\rm acm} - \lambda^2\mu_1^{\rm acm}}{2 M}
\nonumber\\
&& \times
\int_0^1dx \left[{\mathpzc C}^{\rm iu}(\omega_P) - {\mathpzc C}_1^{\rm iu}(\omega_P)\right]\gamma_\mu\gamma_5 \nonumber\\
&+&
\frac{\alpha_{\rm IR}}{6i\pi m_G^2}\frac{1}{M}
\int_0^1dx\,\bar {\mathpzc C}_1^{\rm iu}(\omega_P) \nonumber \\
&&\times
\bigg\{3\mu_-^{\rm acm}\gamma\cdot(p + (x - 1/2)P)\nonumber \\
&& \quad \times [(1 - x)\lambda^1 - x\lambda^2]P_\mu  \nonumber \\
&& + i\big[(1 - x)\lambda^1
+ x\lambda^2\big)] \big[\mu_1^{\rm acm}\gamma_\nu P_\alpha\sigma_{\alpha\mu} \nonumber\\
&& \quad - \mu_2^{\rm acm}P_\alpha\sigma_{\alpha\mu}\gamma_\nu\big](p + (x - 1/2)P)_\nu \nonumber\\
&&- \lambda^-M \big[\mu_+^{\rm acm}(p + (x - 1/2)P)_\beta\sigma_{\mu\beta} \nonumber \\
&& \quad + 3i\mu_-^{\rm acm}(p + (x - 1/2)P)_\mu\big]\bigg\}\,\gamma_5\,, \label{correctionmuacm}
\end{eqnarray}
where $\mu_\pm^{\rm acm} = \mu_1^{\rm acm} \pm \mu_2^{\rm acm}$, and $\{\lambda^i,\,i=1,2\}$ represents the quark propagator correction on each leg with $\lambda^\pm = \lambda^1\pm\lambda^2$.

One can now adapt the general expression in Eq.\,\eqref{correctionmuacm} to the particular cases of relevance herein.  The first is the $\rho$-meson Bethe-Salpeter amplitude. Capitalising on isospin symmetry, which entails $\mu^{\rm acm}_u=\mu^{\rm acm}_d=:\mu^{\rm acm}$, one finds
\begin{eqnarray}
\Gamma_\alpha^{\rho\,{\rm acm}} & = &
\frac{\alpha_{\rm IR}}{2i\pi m_G^2}
\frac{\mu^{\rm acm}\lambda^-}{2 M}
E_\rho\int_0^1dx\left[ {\mathpzc C}^{\rm iu}(\omega_P) - {\mathpzc C}_1^{\rm iu}(\omega_P)\right] \nonumber\\
&& \times \gamma_\mu\mathcal P_{\mu\alpha}\gamma_5 \nonumber\\
&+&\frac{\alpha_{\rm IR}}{3i\pi m_G^2}\frac{\mu^{\rm acm}}{2 M}
E_\rho  \int_0^1dx\,\bar {\mathpzc C}_1^{\rm iu}(\omega_P)
\bigg\{ i\big[(1 - x)\lambda^1  \nonumber\\
&&+ x\lambda^2\big]\big(\gamma_\beta P_\nu\sigma_{\nu\alpha} - P_\nu\sigma_{\nu\alpha}\gamma_\beta\big)(p + (x - 1/2)P)_\beta \nonumber\\
&&- 2\lambda^-M\mathcal P_{\mu\alpha}(p + (x - 1/2)P)_\nu\sigma_{\mu\nu}\bigg\}\,\gamma_5\,,
\end{eqnarray}
where ``$\lambda$'' is constructed from the correction specified in one of Eqs.\,\eqref{lambda1} -- \eqref{lambda3}.

The other case is the quark-photon vertex, for which the correction is found with $\lambda^1 = \lambda^2 = \lambda$, since the quark flavours are identical, and we need only consider $q^2=0$:
\begin{equation}
\Gamma_\mu^{\gamma\,{\rm acm}} =
\frac{\alpha_{\rm IR}}{3\pi m_G^2} \frac{\mu^{\rm acm}\lambda }{2 M}
\bar {\mathpzc C}_1^{\rm iu}(M^2) \gamma_5
\big[\gamma\cdot p\, , \,\sigma_{\mu\alpha}q_\alpha
\big]\,.
\end{equation}

\section{$\mathbf{\rho}$-meson EDM: Results}
\label{sec:results}
\subsection{Analysis without Peccei-Quinn symmetry}
In order to obtain a result for the $\rho$-meson's EDM, $d_\rho$, it remains only to sum the various contributions derived in Sec.\,\ref{sec:formulae} as they contribute to Eq.\,\eqref{GIArho}, evaluated with the parameter values in Table~\ref{Table:static}:
\begin{eqnarray}
d_\rho&=&
%-\frac{2.88\times10^{-3}}{\mathpzc s}\,\mu^{\rm acm}\,e\bar\theta \nonumber\\
- 2.88\times10^{-3}\,\mu^{\rm acm}\,e\bar\theta/\mathpzc s \nonumber\\
&&
+  0.785\,(d_u - d_d) \nonumber\\
&& + (1.352 + 0.775\,\mu^{\rm acm})e(\tilde d_u - \tilde d_d) \nonumber\\
&&  - (0.091 - 2.396\,\mu^{\rm acm}) e(\tilde d_u + \tilde d_d) \nonumber \\
&& - e\frac{{\mathpzc s}{\mathpzc K}}{\Lambda^2} \, (2.696 - 6.798\,\mu^{\rm acm})\times10^{-3}\,.
\label{drhoANS}
\end{eqnarray}
In this formula, $d_f$, $\tilde d_f$ carry a dimension of inverse-mass and ${\mathpzc s} = 1$\,GeV.  %N.B.\ In some analyses, contrary to our usage, a factor of $e$ is included in $d_f$.

A nugatory transformation allows one to rewrite Eq.\,\eqref{drhoANS} in terms of dimensionless electric and chromoelectric quark dipole moments; viz.,
\begin{eqnarray}
d_\rho&=&
%-\frac{2.88\times10^{-3}}{\mathpzc s}\,\mu^{\rm acm}\,e\bar\theta \nonumber\\
- 2.88\times10^{-3}\,\mu^{\rm acm}\,e\bar\theta/\mathpzc s \nonumber\\
&&
+ \frac{\mathpzc{v}_H}{\Lambda^2}
\bigg[
0.785\,(D_u - D_d) \nonumber\\
&& \quad + (1.352 + 0.775\,\mu^{\rm acm})e(\tilde D_u - \tilde D_d) \nonumber\\
&& \quad - (0.091 - 2.396\,\mu^{\rm acm}) e(\tilde D_u + \tilde D_d) \nonumber \\
&& \quad - (1.096 - 2.763\, \mu^{\rm acm})\times10^{-5}\,e\mathpzc K\bigg]\,,
\label{drhoANSN}
\end{eqnarray}
where $\mathpzc{v}_H=246\,$GeV is the cube-root of the phenomenological Higgs vacuum expectation value.  In a class of models that includes, e.g., Ref.\,\cite{Weinberg:1976hu}, one finds
\begin{equation}
D_f \sim \frac{m_f}{\mathpzc{v}_H} \sim 2\times 10^{-5},
\label{Dfguess}
\end{equation}
a result which may be used to inform expectations about the ``natural'' magnitude of the terms in Eqs.\,\eqref{drhoANS}, \eqref{drhoANSN}.

There are four distinct types of contribution to $d_\rho$ in Eq.\,\eqref{drhoANS}.  The first is associated with the $\theta$-term; and it is notable that this contribution vanishes in the absence of a dressed-quark anomalous magnetic moment, a feature which emphasises the connection between topology and dynamical chiral symmetry breaking that is highlighted, e.g., in Eq.\,(21) of Ref.\,\cite{Bhagwat:2007ha}.  Our result may directly be compared with that obtained in a sum rules analysis; viz.,
\begin{equation}
\begin{array}{rl}
\mbox{herein}: &- 2.9\times10^{-3}\,\mu^{\rm acm}\,e\bar\theta \sim
%0.72 \times 10^{-3}\,e\bar\theta \\
0.7 \times 10^{-3}\,e\bar\theta \\
\mbox{\rm Ref.\,\protect\cite{Pospelov:1999rg}}: & \rule{2ex}{0ex} 4.4 \times 10^{-3}\, e\bar\theta \,.
\end{array}
\end{equation}

\begin{figure}[t]
\centerline{%
\includegraphics[clip,width=0.45\textwidth]{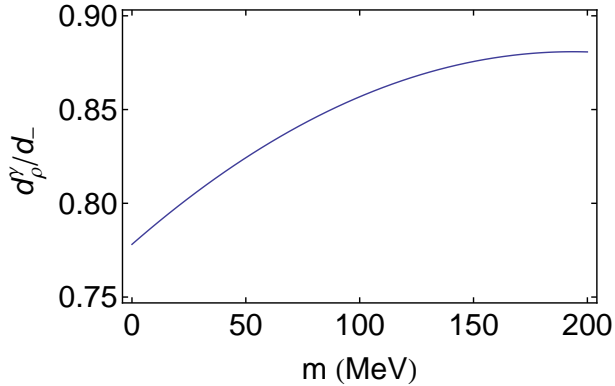}}
\caption{\label{fig:dEDM} Evolution of the quark-EDM component of the $\rho$-meson's EDM with current-quark mass, assuming $d_-$ is independent of $m$.  $m=170\,$MeV corresponds to the mass of the $s$-quark in our treatment of the contact interaction \protect\cite{Chen:2012qr}, so the difference between $d^\gamma_\rho$ and $d^\gamma_\phi$ is 10\%.}
\end{figure}

The second contribution owes to an explicit dressed-quark EDM.  It has been computed via a number of methods, so that a comparison with our results is readily compiled:
\begin{equation}
\begin{array}{ccccc}
\mbox{herein} & \mbox{DSE\,\protect\cite{Hecht:1997uj}} &
\mbox{BM\,\protect\cite{Hecht:1997uj}} &
\mbox{nrQM\,\protect\cite{Hecht:1997uj}} &
\mbox{sum\,rules\,\protect\cite{Pospelov:1999rg}}\\
0.79  & 0.72  & 0.83 & 1.00 & 0.51\,,
\end{array}
\label{drhoEDM}
\end{equation}
where each entry is multiplied by $d_-=(d_u - d_d)$; and
\mbox{DSE}\,\protect\cite{Hecht:1997uj} summarises results obtained from momentum-dependent DSE input,
\mbox{BM}\,\protect\cite{Hecht:1997uj} reports a bag-model result,
and \mbox{nrQM}\,\protect\cite{Hecht:1997uj} is the non-relativistic constituent-quark value.
We depict the current-quark mass dependence of this contribution in Fig.\,\ref{fig:dEDM}.
It is notable that the magnitude of these results matches an existing DSE estimate of the analogous contribution to the neutron's EDM \cite{Hecht:2001ry}.  Moreover, based on Ref.\,\cite{Pallaghy:1996}, a perturbative analysis would yield $2 m_\rho d_\rho^{\rm pert} = 2 m \, d_-$, where $m$ is the current-quark mass.  With the parameter values employed herein, this is $d_\rho^{\rm pert} = 0.014 \,d_-$, which is just $\sim 2$\% of the order-of-magnitude specified by the values in Eq.\,\eqref{drhoEDM}.

\begin{table}[t]
\caption{Contributions to the $\rho$-meson EDM associated with a quark chromoelectric dipole moment, with $\tilde d^e_\mp = e(\tilde d_u \mp \tilde d_d)$.
Row~1: quark-photon vertex correction, Sec.\,\protect\ref{sss:LCEDMQPV}; Row~2: $\rho$-meson Bethe-Salpeter amplitude correction, Sec.\,\protect\ref{sss:LCEDMBSA}; Row~3: dressed-quark propagator correction, Sec.\,\protect\ref{sss:LCEDMS}; Row~4: anomalous chromomagnetic moment contributions, Sec.\,\protect\ref{sec:ACM}; Row~5: sum of preceding four rows;
Row~6: Row~5 evaluated with $\mu^{\rm acm}=-1/4$; and Row~7: sum rules result from Ref.\,\protect\cite{Pospelov:1999rg}, evaluated here with a heavy $s$-quark.
\label{tablegCEDM}
}
\begin{center}
\begin{tabular*}%{|c|c|c|c|c|c|c|}\hline
{\hsize}
{
l@{\extracolsep{0ptplus1fil}}
|c@{\extracolsep{0ptplus1fil}}}\hline
$q\gamma q$ \rule{1em}{0ex} & $-0.066\,\tilde d^e_- - 0.199\,\tilde d^e_+$\\
BSA & $-0.120\,\tilde d^e_- + 0.108\,\tilde d^e_+$\\
$S(k)$ & $1.538\,\tilde d^e_-  \phantom{+0.108\,\tilde d^e_+}$ \\
acm $(\times \mu^{\rm acm})$ & $\phantom{-} 0.775\,\tilde d^e_-  + 2.396\,\tilde d^e_+ $\\\hline
\emph{our CEDM} & $(1.35 + 0.78\,\mu^{\rm acm})\,\tilde d^e_-
- (0.09 - 2.40\,\mu^{\rm acm})\,\tilde d^e_+$ \\
 \emph{total} &   $\phantom{-}1.16 \,\tilde d^e_- - 0.69 \,\tilde d^e_+$ \\\hline
sum rules \cite{Pospelov:1999rg} & $-0.13\,\tilde d^e_- \phantom{- 0.58\,\tilde d_+}$
%...PQ symmetry follows
%sum rules \cite{Pospelov:1999rg} & $-0.34\,\tilde d_-  - 0.58\,\tilde d_+ $
%
\\ \hline
\end{tabular*}
\end{center}
\end{table}

The third contribution to $d_\rho$ is generated by the quark's chromoelectric dipole moment.  Its subcomponents are detailed in Table~\ref{tablegCEDM}.  The net result is comparable in magnitude and sign with that produced by the quark EDM, Eq.\,\eqref{drhoEDM}.  In comparison with a sum rules computation \cite{Pospelov:1999rg}, however, our result is an order of magnitude larger, has the opposite sign and contains a sizeable $\tilde d_+$-term.  At least the first two of these marked discrepancies are insensitive to reasonable variations in $\mu^{\rm acm}$.  It is worth emphasising here that our calculation has no other variable parameters: the two specifying our model, listed in Table~\ref{Table:static}, were fixed in prior studies of an array of meson and baryon observables \cite{GutierrezGuerrero:2010md,Roberts:2010rn,Roberts:2011cf,%
Roberts:2011wy,Wilson:2011aa,Chen:2012qr}.
This mismatch will receive further attention in future work.

% In particular because these authors can't agree on what to write now.
%
%A context for the mismatch between our result for the quark chromo-EDM contribution to $d_\rho$ and that described in Ref.\,\cite{Pospelov:1999rg} is provided by the following observations.  In connection with estimates of the neutron's EDM, the sum rules result for the quark chromo-EDM contribution is twenty-times smaller than that obtained within the context of chiral effective field theory \cite{Khatsymovsky:1992yg}.
%
%It is also relevant to note here that modern analyses show the strangeness content of light-quark hadrons to be very small \cite{Young:2006jc,Cloet:2008fw,Young:2009zb,Chang:2009ae,Durr:2011mp,Horsley:2011wr}.
%
%It is conceivable, therefore, that sum rules are not well suited to estimating the quark chromo-EDM contribution to an hadron's EDM.

The four-fermion interaction is responsible for the final contribution in Eq.\,\eqref{drhoANS}.  Its subcomponents are detailed in Table~\ref{tableD6}.  As ours is the first estimate of the contribution from a dimension-six operator to the $\rho$-meson's EDM, there is no ready substantial comparison.  On the other hand, the result in Table~\ref{tableD6} is quickly seen to be ``natural'' in size.
The dimension-six operator is associated with a coupling ${\mathpzc K} /\Lambda^2$,
which has mass-dimension ``$-2$''.  In order to obtain a quantity with mass-dimension ``$-1$'', this coupling must be multiplied by another energy scale.  We are interested in an hadronic EDM, so that scale should be typical of hadron physics; e.g., the dressed-quark mass ``$M$''.  Finally, a loop correction is required for the generation of an EDM, and loops are characterised by a factor $1/(16\pi^2)$.  Putting these quantities together yields an expectation based on naive dimensional analysis; viz.,
\begin{equation}
d_\rho^{D=6} \sim e\frac{1}{16\pi^2} \, \frac{M}{{\mathpzc v}_H } \frac{{\mathpzc v}_H {\mathpzc K}}{\Lambda^2} \sim 1\times 10^{-5}\frac{e{\mathpzc v}_H {\mathpzc K}}{{\Lambda}^2}\,,
\end{equation}
in agreement with the magnitude of the final row in Table~\ref{tableD6}.  Comparison with Eq.\,\eqref{Dfguess}, furthermore, indicates that in our computation the quark-EDM and dimension-six contributions are naturally related via
\begin{equation}
d_\rho^{q{\rm EDM}}{\mathpzc K} \sim d_\rho^{D=6}.
\end{equation}

\begin{table}[t]
\caption{Contributions to the $\rho$-meson EDM associated with the dimension-six operator in Eq.\,\protect\eqref{LeffCompute}.  Each row should be multiplied by $ e{\mathpzc v}_H {\mathpzc K} /{\Lambda}^2$.
Row~1: quark-photon vertex correction, Sec.\,\protect\ref{sss:L6QPV}; Row~2: $\rho$-meson Bethe-Salpeter amplitude correction, Sec.\,\protect\ref{sss:L6BSA}; Row~3: dressed-quark propagator correction, Sec.\,\protect\ref{sss:L6S}; Row~4: anomalous chromomagnetic moment contributions, Sec.\,\protect\ref{sec:ACM}; and Row~5: sum of preceding four rows.
\label{tableD6}
}
\begin{center}
\begin{tabular*}%{|c|c|c|c|c|c|c|}\hline
{\hsize}
{
l@{\extracolsep{0ptplus1fil}}
|c@{\extracolsep{0ptplus1fil}}}\hline
$q\gamma q$ \rule{1em}{0ex} & $ - 1.005 \times 10^{-5} $\\
BSA & $- 9.114 \times 10^{-7}$\\
$S(k)$ & $0$ \\
acm $(\times \mu^{\rm acm})$ & $2.763 \times 10^{-5}\,\mu^{\rm acm} \phantom{\,\mu^{\rm acm}}$\\\hline
\emph{our $D=6$ total} & $-(1.096 -2.763\, \mu^{\rm acm}) \times 10^{-5}$ \\\hline
\end{tabular*}
\end{center}
\end{table}

\subsection{Peccei Quinn Symmetry}
The leading term in Eq.\,\eqref{drhoANSN} is that associated with $\bar\theta$.  Arising from a dimension-four operator, this contribution is not suppressed by a large beyond-SM mass-scale.  One may furthermore expect that, absent any symmetry to prevent it, a typical non-SM for CP-violation will produce large corrections to $\bar\theta$.  In order to reconcile this with the remarkably small upper-bound on $\bar\theta$ placed by the neutron's EDM, one must accept that the initial value of $\bar\theta$ is very finely tuned.  There is nothing to prevent this from being simply an accident of Nature.  However, some view that possibility as aesthetically displeasing and prefer to introduce a new dynamical degree of freedom, the axion, a pseudo-Goldstone boson, whose role is to cancel the effect of $\bar\theta$ \cite{Peccei:1977hh}.  It is notable that there is currently no empirical evidence in favour of the axion's existence and the remaining domain of parameter space is small \cite{Kim:2009xp}.

Notwithstanding this, in the context of EDM estimates it is customary to expose the possible effect of axion physics on the results in Eq.\,\eqref{drhoANS} or \eqref{drhoANSN}.  Here there is a complication.  If one considers an extension of the SM with a collection of CP-odd operators that may mix with the $\bar\theta$-term, then the effective potential describing axion physics at the hadronic scale can plausibly acquire terms that shift its minimum to a nonzero value of the effective $\bar\theta$-parameter, $\bar\theta_{\rm induced}$ \cite{Pospelov:2005pr}.  The quark chromoelectric dipole moment interaction is one such operator.  In its case, within a sum rules calculation \cite{Pospelov:1999rg}, the net effect of this mixing is elimination of $\bar\theta$ in favour of a modest enhancement in magnitude of the coefficients of $\tilde d_\pm$ in Eq.\,\eqref{drhoANS}, with no change in sign.

The implications for our study are plain.  Allowing an axion-like mechanism to play a role, then $\bar\theta$ disappears from Eqs.\,\eqref{drhoANS} and \eqref{drhoANSN}, and any measurement of an hadron EDM, here that of the $\rho$-meson, places a little more stringent constraint on $\tilde d_\pm$ in particular but also on $d_\pm$ and $\mathpzc K$.

%This is, perhaps, particularly relevant to $\mathpzc K$, since, in some supersymmetric (SUSY) extensions of the SM, the associated term in Eq.\,\eqref{LeffCompute} can be generated via a one-loop SUSY correction to the quark propagator.  Such a correction can be shifted into a complex phase in the current-quark mass matrix, so constraints on $\bar\theta$ place a bound on $\mathpzc K$.  On the other hand, with the elimination of $\bar\theta$ via an axion effective potential, the term modulated by $\mathpzc K$ is exposed to independent constraint \cite{RamseyMusolf:2006vr}.

This is, perhaps, particularly relevant to $\mathcal{K}$, since the high-scale physics that generates this operator will typically also produce a complex phase for the quark masses.  Within the low-energy effective theory of Eq.\,\eqref{LeffCompute}, this phase will arise from one-loop contributions to the quark propagator containing one insertion of the CP-violating four-quark operator and the quark Yukawa interaction.  Consequently, constraints on ${\bar\theta}$ imply a bound on $\mathcal{K}$.  On the other hand, with the elimination of $\bar\theta$ via an axion effective potential, the term modulated by $\mathpzc K$ is exposed to independent constraint \cite{RamseyMusolf:2006vr}.
Computing the contribution of the four-quark CP-violating operator to the axion potential, determining the resulting dependence of ${\bar\theta}_\mathrm{induced}$ on $\mathcal{K}$, and deriving the expression corresponding to Eq.\,\eqref{drhoANSN} will be the subject of future work.

\section{Epilogue}
\label{sec:epilogue}
Using the leading-order in a global-symmetry-preserving truncation of QCD's Dyson-Schwinger equations, we computed the electric dipole moment of the $\rho$-meson, $d_\rho$, that is generated by the leading dimension-four and -five CP-violating operators and an example of a dimension-six operator.  We employed a momentum-independent form for the leading-order kernel in the gap- and Bethe-Salpeter equations.  This is known to produce results for low-energy pseudoscalar- and vector-meson observables that are indistinguishable from those obtained with the most sophisticated interactions available when they are analysed using the same truncation.  Since the dipole moment is a low-energy observable, our predictions should be similarly reliable, in which case the framework we employ and elucidate can usefully be adapted to the more challenging task of computing the neutron's EDM, $d_n$.

We find that the two dimension-five operators; namely, quark-EDM and -chromo-EDM, characterised by $d_q$ and $\tilde d_q$, respectively, produce contributions to $d_\rho$ whose coefficients are of the same sign and within a factor of two in magnitude.  This contrasts with an extant sum rules evaluation, in which the coefficients of the contributions have the opposite sign and differ by a factor of four in magnitude.  Since all studies agree within a factor of two on the quark-EDM coefficient, the discrepancy resides with the chromo-EDM contribution.  %This is notable because a sum rules evaluation of the quark chromo-EDM contribution to $d_n$ is more than an order-of-magnitude smaller than a result obtained in a study based on chiral effective theory.
These differences invite further analysis and guarantee relevance to a DSE evaluation of the impact of $\tilde d_q$ on the neutron's EDM.

Absent a mechanism that suppresses a $\theta$-term in any beyond-Standard-Model action, the tight constraints on the magnitude of a contribution from this term to the neutron's EDM also apply to contributions from a dimension-six four-fermion operator to this or another hadron's EDM.  Should such a mechanism exist, however, we find that a dimension-six operator can match the quark-EDM and chromo-EDM in importance.

Using the techniques described herein, calculation of the neutron's EDM is underway.

\begin{acknowledgments}
We thank A.~Bashir, L.~Chang, C.~Chen and B.\,H.\,J.~McKellar
for helpful comments.
This work was supported in part by:
U.\,S.\ Department of Energy, Office of Nuclear Physics, contract nos.~DE-AC02-06CH11357 and DE-FG02-08ER41531; Forschungszentrum J\"ulich GmbH; and the Wisconsin Alumni
Research Foundation.
\end{acknowledgments}

\bibliographystyle{../../../../../zProc/z10KITPC/h-physrev4}
\bibliography{../../../../../CollectedBiB}

\begin{thebibliography}{10}

\bibitem{Purcell:1950zz}
E.~Purcell and N.~Ramsey,
\newblock Phys. Rev. {\bf 78}, 807 (1950).
%%CITATION = PHRVA,78,807;%%

\bibitem{Sakharov:1967dj}
A.~D. Sakharov,
\newblock Pisma Zh. Eksp. Teor. Fiz. {\bf 5}, 32 (1967).
%%CITATION = ZFPRA,5,32;%%

\bibitem{Cohen:1993nk}
A.~G. Cohen, D.~B. Kaplan and A.~E. Nelson,
\newblock Ann. Rev. Nucl. Part. Sci. {\bf 43}, 27 (1993).

\bibitem{Trodden:1998ym}
M.~Trodden,
\newblock Rev.Mod.Phys. {\bf 71}, 1463 (1999).

\bibitem{Morrissey:2012db}
D.~E. Morrissey and M.~J. Ramsey-Musolf,
\newblock arXiv:1206.2942{\,}[hep-ph],
\newblock {\mbox{\emph{Electroweak baryogenesis}}}.
%%CITATION = ARXIV:1206.2942;%%

\bibitem{Harris:1999jx}
P.~Harris {\em et~al.},
\newblock Phys. Rev. Lett. {\bf 82}, 904 (1999).
%%CITATION = PRLTA,82,904;%%

\bibitem{Pospelov:2005pr}
M.~Pospelov and A.~Ritz,
\newblock Annals Phys. {\bf 318}, 119 (2005).

\bibitem{RamseyMusolf:2006vr}
M.~Ramsey-Musolf and S.~Su,
\newblock Phys.Rept. {\bf 456}, 1 (2008).

\bibitem{Lamoreaux:2009zz}
S.~Lamoreaux and R.~Golub,
\newblock J. Phys. G {\bf G36}, 104002 (2009).
%%CITATION = JPHGB,G36,104002;%%

\bibitem{Cirigliano:2009yd}
V.~Cirigliano, Y.~Li, S.~Profumo and M.~J. Ramsey-Musolf,
\newblock JHEP {\bf 1001}, 002 (2010).
%%CITATION = ARXIV:0910.4589;%%

\bibitem{Li:2010ax}
Y.~Li, S.~Profumo and M.~Ramsey-Musolf,
\newblock JHEP {\bf 1008}, 062 (2010).
%%CITATION = ARXIV:1006.1440;%%

\bibitem{Kozaczuk:2012xv}
J.~Kozaczuk, S.~Profumo, M.~J. Ramsey-Musolf and C.~L. Wainwright,
\newblock arXiv:1206.4100 [hep-ph],
\newblock {\emph{Supersymmetric Electroweak Baryogenesis Via Resonant Sfermion
  Sources}}.
%%CITATION = ARXIV:1206.4100;%%

\bibitem{Hecht:1997uj}
M.~B. Hecht and B.~H.~J. McKellar,
\newblock Phys. Rev. {\bf C57}, 2638 (1998).

\bibitem{Hecht:1999fd}
M.~B. Hecht and B.~H.~J. McKellar,
\newblock Phys. Rev. {\bf C60}, 065202 (1999).
%%CITATION = HEP-PH/9906246;%%

\bibitem{Pospelov:1999rg}
M.~Pospelov and A.~Ritz,
\newblock Phys. Lett. {\bf B471}, 388 (2000).

\bibitem{Weinberg:1989dx}
S.~Weinberg,
\newblock Phys.Rev.Lett. {\bf 63}, 2333 (1989).
%%CITATION = PRLTA,63,2333;%%

\bibitem{GutierrezGuerrero:2010md}
L.~X. Guti{\'e}rrez-Guerrero, A.~Bashir, I.~C. Clo{\"e}t and C.~D. Roberts,
\newblock Phys. Rev. {\bf C81}, 065202 (2010).
%%CITATION = 1002.1968;%%

\bibitem{Roberts:2010rn}
H.~L.~L. Roberts, C.~D. Roberts, A.~Bashir, L.~X. Guti{\'e}rrez-Guerrero and
  P.~C. Tandy,
\newblock Phys. Rev. {\bf C82}, 065202 (2010).
%%CITATION = 1009.0067;%%

\bibitem{Roberts:2011cf}
H.~L.~L. Roberts, L.~Chang, I.~C. Clo{\"e}t and C.~D. Roberts,
\newblock Few Body Syst. {\bf 51}, 1 (2011).
%%CITATION = 1101.4244;%%

\bibitem{Roberts:2011wy}
H.~L.~L. Roberts, A.~Bashir, L.~X. Guti{\'e}rrez-Guerrero, C.~D. Roberts and
  D.~J. Wilson,
\newblock Phys. Rev. {\bf C83}, 065206 (2011).
%%CITATION = 1102.4376;%%

\bibitem{Wilson:2011aa}
D.~J. Wilson, I.~C. Clo{\"e}t, L.~Chang and C.~D. Roberts,
\newblock Phys. Rev. {\bf C85}, 025205 (2012).
%%CITATION = 1112.2212;%%

\bibitem{Chen:2012qr}
C.~Chen, L.~Chang, C.~D. Roberts, S.~Wan and D.~J. Wilson,
\newblock Few Body Syst. \emph{in press}  (2012), [arXiv:1204.2553 nucl-th].
%%CITATION = ARXIV:1204.2553;%%

\bibitem{Chang:2011vu}
L.~Chang, C.~D. Roberts and P.~C. Tandy,
\newblock Chin. J. Phys. {\bf 49}, 955 (2011).
%%CITATION = 1107.4003;%%

\bibitem{Bashir:2012fs}
A.~Bashir {\em et~al.},
\newblock Commun. Theor. Phys. {\bf 58}, 79 (2012).
%%CITATION = ARXIV:1201.3366;%%

\bibitem{Roberts:2012sv}
C.~D. Roberts,
\newblock arXiv:1203.5341 [nucl-th],
\newblock {{\emph{Strong QCD and Dyson-Schwinger Equations}}}.
%%CITATION = ARXIV:1203.5341;%%

\bibitem{Jain:1993qh}
P.~Jain and H.~J. Munczek,
\newblock Phys. Rev. {\bf D48}, 5403 (1993).
%%CITATION = HEP-PH/9307221;%%

\bibitem{Maris:1997tm}
P.~Maris and C.~D. Roberts,
\newblock Phys. Rev. {\bf C56}, 3369 (1997).
%%CITATION = NUCL-TH/9708029;%%

\bibitem{Maris:2003vk}
P.~Maris and C.~D. Roberts,
\newblock Int. J. Mod. Phys. {\bf E12}, 297 (2003).
%%CITATION = NUCL-TH/0301049;%%

\bibitem{Munczek:1994zz}
H.~J. Munczek,
\newblock Phys. Rev. {\bf D52}, 4736 (1995).
%%CITATION = HEP-TH/9411239;%%

\bibitem{Bender:1996bb}
A.~Bender, C.~D. Roberts and L.~von Smekal,
\newblock Phys. Lett. {\bf B380}, 7 (1996).
%%CITATION = NUCL-TH/9602012;%%

\bibitem{Maris:2002mt}
P.~Maris, A.~Raya, C.~D. Roberts and S.~M. Schmidt,
\newblock Eur. Phys. J. {\bf A18}, 231 (2003).
%%CITATION = NUCL-TH/0208071;%%

\bibitem{Eichmann:2008ae}
G.~Eichmann, R.~Alkofer, I.~C. Clo{\"e}t, A.~Krassnigg and C.~D. Roberts,
\newblock Phys. Rev. {\bf C77}, 042202(R) (2008).
%%CITATION = 0802.1948;%%

\bibitem{Maris:1999nt}
P.~Maris and P.~C. Tandy,
\newblock Phys. Rev. {\bf C60}, 055214 (1999).
%%CITATION = NUCL-TH/9905056;%%

\bibitem{Bhagwat:2006pu}
M.~S. Bhagwat and P.~Maris,
\newblock Phys. Rev. {\bf C77}, 025203 (2008).
%%CITATION = NUCL-TH/0612069;%%

\bibitem{Qin:2011dd}
S.-x. Qin, L.~Chang, Y.-x. Liu, C.~D. Roberts and D.~J. Wilson,
\newblock Phys. Rev. {\bf C84}, 042202(R) (2011).
%%CITATION = 1108.0603;%%

\bibitem{Qin:2011xq}
S.-x. Qin, L.~Chang, Y.-x. Liu, C.~D. Roberts and D.~J. Wilson,
\newblock Phys. Rev. {\bf C85}, 035202 (2012).
%%CITATION = 1109.3459;%%

\bibitem{Maris:1998hc}
P.~Maris and C.~D. Roberts,
\newblock Phys. Rev. {\bf C58}, 3659 (1998).
%%CITATION = NUCL-TH/9804062;%%

\bibitem{Maris:2000sk}
P.~Maris and P.~C. Tandy,
\newblock Phys. Rev. {\bf C62}, 055204 (2000).
%%CITATION = NUCL-TH/0005015;%%

\bibitem{Maris:2002mz}
P.~Maris and P.~C. Tandy,
\newblock Phys. Rev. {\bf C65}, 045211 (2002).
%%CITATION = NUCL-TH/0201017;%%

\bibitem{Bhagwat:2007ha}
M.~S. Bhagwat, L.~Chang, Y.-X. Liu, C.~D. Roberts and P.~C. Tandy,
\newblock Phys. Rev. {\bf C76}, 045203 (2007).
%%CITATION = 0708.1118;%%

\bibitem{Eichmann:2008ef}
G.~Eichmann, I.~C. Clo{\"e}t, R.~Alkofer, A.~Krassnigg and C.~D. Roberts,
\newblock Phys. Rev. {\bf C79}, 012202 (2009).
%%CITATION = 0810.1222;%%

\bibitem{Eichmann:2011vu}
G.~Eichmann,
\newblock Phys. Rev. {\bf D84}, 014014 (2011).
%%CITATION = 1104.4505;%%

\bibitem{Eichmann:2011ej}
G.~Eichmann,
\newblock PoS {\bf QCD-TNT-II}, 017 (2011).
%%CITATION = ARXIV:1112.4888;%%

\bibitem{Eichmann:2011pv}
G.~Eichmann and C.~Fischer,
\newblock Eur. Phys. J. {\bf A48}, 9 (2012).
%%CITATION = ARXIV:1111.2614;%%

\bibitem{Bhagwat:2006xi}
M.~S. Bhagwat, A.~Krassnigg, P.~Maris and C.~D. Roberts,
\newblock Eur. Phys. J. {\bf A31}, 630 (2007).
%%CITATION = NUCL-TH/0612027;%%

\bibitem{Aguilar:2010gm}
A.~C. Aguilar, D.~Binosi and J.~Papavassiliou,
\newblock JHEP {\bf 07}, 002 (2010).
%%CITATION = 1004.1105;%%

\bibitem{Boucaud:2010gr}
P.~Boucaud {\em et~al.},
\newblock Phys. Rev. {\bf D82}, 054007 (2010).
%%CITATION = ARXIV:1004.4135;%%

\bibitem{Nambu:1961tp}
Y.~Nambu and G.~Jona-Lasinio,
\newblock Phys. Rev. {\bf 122}, 345 (1961).
%%CITATION = PHRVA,122,345;%%

\bibitem{Ebert:1996vx}
D.~Ebert, T.~Feldmann and H.~Reinhardt,
\newblock Phys. Lett. {\bf B388}, 154 (1996).
%%CITATION = HEP-PH/9608223;%%

\bibitem{Krein:1990sf}
C.~D. Roberts, A.~G. Williams and G.~Krein,
\newblock Int. J. Mod. Phys. {\bf A7}, 5607 (1992).
%%CITATION = IMPAE,A7,5607;%%

\bibitem{Brodsky:2010xf}
S.~J. Brodsky, C.~D. Roberts, R.~Shrock and P.~C. Tandy,
\newblock Phys. Rev. {\bf C82}, 022201(R) (2010).
%%CITATION = 1005.4610;%%

\bibitem{Chang:2011mu}
L.~Chang, C.~D. Roberts and P.~C. Tandy,
\newblock Phys. Rev. {\bf C85}, 012201(R) (2012).
%%CITATION = 1109.2903;%%

\bibitem{Brodsky:2012ku}
S.~J. Brodsky, C.~D. Roberts, R.~Shrock and P.~C. Tandy,
\newblock Phys. Rev. {\bf C{\,}85}, 065202 (2012).
%%CITATION = ARXIV:1202.2376;%%

\bibitem{Nakamura:2010zzi}
K.~Nakamura {\em et~al.},
\newblock J. Phys. {\bf G37}, 075021 (2010).
%%CITATION = JPHGB,G37,075021;%%

\bibitem{Chang:2010hb}
L.~Chang, Y.-X. Liu and C.~D. Roberts,
\newblock Phys. Rev. Lett. {\bf 106}, 072001 (2011).
%%CITATION = 1009.3458;%%

\bibitem{Hawes:1998bz}
F.~T. Hawes and M.~A. Pichowsky,
\newblock Phys. Rev. {\bf C59}, 1743 (1999).

\bibitem{deMelo:1997hh}
J.~de~Melo and T.~Frederico,
\newblock Phys. Rev. {\bf C55}, 2043 (1997).
%%CITATION = NUCL-TH/9706032;%%

\bibitem{Choi:2004ww}
H.-M. Choi and C.-R. Ji,
\newblock Phys. Rev. {\bf D70}, 053015 (2004).
%%CITATION = HEP-PH/0402114;%%

\bibitem{Samsonov:2003hs}
A.~Samsonov,
\newblock JHEP {\bf 0312}, 061 (2003).
%%CITATION = HEP-PH/0308065;%%

\bibitem{Bhagwat:2003vw}
M.~Bhagwat, M.~Pichowsky, C.~Roberts and P.~Tandy,
\newblock Phys. Rev. {\bf C68}, 015203 (2003).
%%CITATION = NUCL-TH/0304003;%%

\bibitem{Bowman:2005vx}
P.~O. Bowman {\em et~al.},
\newblock Phys. Rev. {\bf D71}, 054507 (2005).
%%CITATION = HEP-LAT/0501019;%%

\bibitem{Bhagwat:2006tu}
M.~S. Bhagwat and P.~C. Tandy,
\newblock AIP Conf. Proc. {\bf 842}, 225 (2006).
%%CITATION = NUCL-TH/0601020;%%

\bibitem{Bhagwat:2007vx}
M.~S. Bhagwat, I.~C. Clo{\"e}t and C.~D. Roberts,
\newblock (arXiv:0710.2059 [nucl-th]),
\newblock {in \emph{Proceedings of the Workshop on Exclusive Reactions at High
  Momentum Transfer}, Newport News, Virginia, 21-24 May 2007, Eds.\
  A.~Radyushkin and P.~Stoler (World Scientific, Singapore, 2007)}.
%%CITATION = 0710.2059;%%

\bibitem{Chang:2011ei}
L.~Chang and C.~D. Roberts,
\newblock Phys. Rev. {\bf C85}, 052201(R) (2012).
%%CITATION = 1104.4821;%%

\bibitem{Weinberg:1976hu}
S.~Weinberg,
\newblock Phys. Rev. Lett. {\bf 37}, 657 (1976).
%%CITATION = PRLTA,37,657;%%

\bibitem{Hecht:2001ry}
M.~B. Hecht, C.~D. Roberts and S.~M. Schmidt,
\newblock Phys. Rev. {\bf C64}, 025204 (2001).
%%CITATION = NUCL-TH/0101058;%%

\bibitem{Pallaghy:1996}
P.~K. Pallaghy,
\newblock {\em CP violation},
\newblock PhD thesis, University of Melbourne, 1996.

\bibitem{Peccei:1977hh}
R.~Peccei and H.~R. Quinn,
\newblock Phys. Rev. Lett. {\bf 38}, 1440 (1977).
%%CITATION = PRLTA,38,1440;%%

\bibitem{Kim:2009xp}
J.~E. Kim,
\newblock AIP Conf. Proc. {\bf 1200}, 83 (2010).
%%CITATION = ARXIV:0909.3908;%%

\end{thebibliography}

\end{document}